\documentclass[12pt]{article}\usepackage[hyperfootnotes=false]{hyperref}
\usepackage{epsfig}
\usepackage{float}
\usepackage{dsfont}

\usepackage{caption}
\usepackage{subcaption}

\usepackage{amsmath}
\usepackage{amssymb}
\usepackage{graphicx}
\newlength{\abstractwidth}
\renewcommand{\thefootnote}{\fnsymbol{footnote}}
\renewcommand{\thanks}[1]{\footnote{#1}} 
\newcommand{\starttext}{
\setcounter{footnote}{0}
\renewcommand{\thefootnote}{\arabic{footnote}}}

\newcommand{\be}{\begin{equation}}
\newcommand{\bea}{\begin{eqnarray}}
\newcommand{\eea}{\end{eqnarray}}
\newcommand{\beq}{\begin{equation}}
\newcommand{\ee}{\end{equation}}

\def\simleq{\; \raise0.3ex\hbox{$<$\kern-0.75em
\raise-1.1ex\hbox{$\sim$}}\; }
\def\simgeq{\; \raise0.3ex\hbox{$>$\kern-0.75em
\raise-1.1ex\hbox{$\sim$}}\; }

\newcommand\To{\rule{0pt}{4.5ex}}       
\newcommand\Bo{\rule[-3.0ex]{0pt}{0pt}} 

\def\bi{\begin{itemize}}
\def\ei{\end{itemize}}

\def\sc{\setcounter{equation}{0}}

\def\CC{{\cal{C}}}

\def\CH{{\cal{H}}}

\def\bn{\bigskip \noindent}

\def\suk{SU(2^K)}

\def\lsim{ \lower .75ex \hbox{$\sim$} \llap{\raise .27ex
\hbox{$<$}} }
\def\gsim{ \lower .75ex \hbox{$\sim$} \llap{\raise .27ex
\hbox{$>$}} }

\makeatletter
\g@addto@macro\normalsize{%
  \setlength\abovedisplayskip{10pt}
  \setlength\belowdisplayskip{20pt}
  \setlength\abovedisplayshortskip{10pt}
  \setlength\belowdisplayshortskip{20pt}
}
\makeatother

\usepackage{color}

\renewcommand{\title}[1]{\vbox{\center\LARGE{#1}}\vspace{5mm}}
\renewcommand{\author}[1]{\vbox{\center#1}\vspace{5mm}}
\newcommand{\address}[1]{\vbox{\center\em#1}}

\begin{document}
  
\begin{titlepage}

\rightline{}
\bigskip
\bigskip\bigskip\bigskip\bigskip
\bigskip

\centerline{\Large \bf {Quantum Complexity and Negative Curvature }}

\bn

\bigskip
\begin{center}

\author{Adam R. Brown, Leonard Susskind, and Ying Zhao}

\address{Stanford Institute for Theoretical Physics and Department of Physics, \\
Stanford University, Stanford, CA 94305-4060, USA}

\bigskip

\end{center}

\begin{abstract}

  As time passes, once simple quantum states tend to become more complex. For strongly coupled $k$-local Hamiltonians, this growth of computational complexity has been conjectured to follow a distinctive and universal pattern. In this paper we show that the same pattern is exhibited by a much simpler system---classical geodesics on a compact two-dimensional geometry of uniform negative curvature. This striking parallel persists whether the system is allowed to evolve naturally or is perturbed from the outside.

\medskip
\noindent
\end{abstract}

\vfill

\end{titlepage}

\vfill\eject

\tableofcontents

\vfill\eject

\starttext \baselineskip=17.63pt \setcounter{footnote}{0}

\sc
\section{The Geometry of Computation}

There is evidence that the evolution of quantum complexity and the growth of the  geometry behind  black hole horizons  follow identical 
patterns 
\cite{Susskind:2014rva,Stanford:2014jda,Roberts:2014isa}. This is true whether the systems evolve in isolation or are subject to external perturbations. In this paper we will show that a third system---the `analog model', namely a classical nonrelativistic  particle that moves along the geodesics of a 
two-dimensional, compact, negatively curved surface of high genus---shares the same behavior. 
Although we do not fully understand the reasons for this correspondence it most likely has its roots in Nielsen's geometrized approach to complexity \cite{2005quant.ph..2070N,2007quant.ph..1004D,2006Sci...311.1133N}.

\sc
\section{The Evolution of Quantum Complexity} \label{sec2}

In this section we will review the evolution of the quantum complexity of strongly-coupled quantum systems, and highlight some signature phenomena. (We will also briefly review the evolution of classical wormholes, to which quantum complexity has been conjectured to be holographically dual.) In  Sec.~\ref{sec3} we will show that much of the highlighted phenomenology  is reproduced by our simple analog model. 

\subsection*{Fast Scramblers}
\label{fastscramblers}

The systems whose complexity we will be interested in modeling are the fast scramblers. Fast-scramblers are systems that spread the effects of localized disturbances over all the degrees of freedom in a  time logarithmic in the entropy. An example of a fast scrambler made with fermionic qubits is the Sachdev-Ye-Kitaev system \cite{KitaevModel,Sachdev:1992fk,Polchinski:2016xgd,Maldacena:2016hyu}; an example with conventional commuting qubits is the high-temperature phase of the Hamiltonian
\be 
H = \sum_{i_1< i_2<...<i_k}
J_{i_1, i_2,...,i_k}
\sigma_{i_1}\sigma_{i_2}...\sigma_{i_k}.
\label{eq:$k$-local}
\ee
(The $J$s are a set of numerical coefficients, possibly chosen randomly, but centered around a value $J.$ The $\sigma_i$ are single qubit traceless Pauli operators for the $i^\textrm{th}$ qubit. We have suppressed the index structure associated with the Pauli operators.) 

This Hamiltonian is not local, since every qubit couples to every other qubit, but it is  ``$k$-local'', since no term in the Hamiltonian couples together more than $k$ qubits.  It is assumed that the total number of qubits, $K$, is much larger than $k.$ For definiteness we will usually take $k=2.$

The fast scrambler of Eq.~\ref{eq:$k$-local} is a continuous-time Hamiltonian.  We will will also be interested in systems for which time is discrete, for example quantum circuits. A quantum circuit starts with a collection of $K$ qubits and makes them interact via $k$-qubit quantum gates. Random $k$-local quantum circuits are believed to be fast scramblers. 

\begin{figure}[H]
\begin{center}
\includegraphics[scale=1.5]{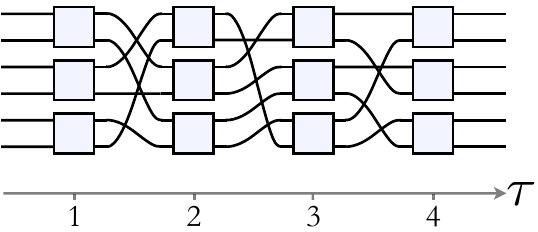}
\caption{An example of a random circuit with $K= 6$, $k=2$, and depth 4. The  six qubits (black lines) are randomly grouped into three ordered pairs, and then a gate (blue box) is  applied to each pair. At the next time-step, they are randomly re-paired.}
\label{circuitarchitecture}
\end{center}
\end{figure}

For $k=2$, a random quantum circuit may be constructed as follows. In each time-step the  $K$ qubits are randomly paired and each pair interacts by a randomly chosen gate. The particular gate set is not very important  as  long as it is universal. After each step the qubits are randomly re-grouped into pairs and the process is repeated. This is illustrated in Fig.~\ref{circuitarchitecture}.
Our focus will not be on the state of the qubit system, but rather on the unitary operator $U(\tau)$ generated by the circuit after $\tau$ time-steps. We will be interested in the complexity of this unitary operator, defined as the number of gates in the minimal quantum circuit that generates this unitary. We will particularly be interested in how the complexity evolves with time. 

\subsection*{Growth and Saturation}
\label{growthandsaturation}

The number of gates that  the circuit applied in order to prepare  $U(\tau)$ is 
\be 
N_\textrm{gates}(\tau) =\frac{K \tau}{2}.
\label{Ngates}
\ee
The number of time-steps, $\tau$, is called the depth of the circuit and $K$ is called the width. The factor of $1/2$ in Eq.~\ref{Ngates} is due to the pairing of qubits, which implies that  in each time-step   $K/2$ gates act.

However, the definition of the complexity $\CC[U(\tau)]$ is not the actual number of gates used  in the defining circuit  to generate $U(\tau).$ Instead the complexity  is
the \it minimum \rm number of gates that it takes to prepare  $U(\tau)$, using the most efficient possible circuit. This may be less than or equal to the actual number of gates used in the defining circuit. However it is believed that at least for some length of time the  defining circuit is the most efficient, and for this period the complexity grows as $\frac{K\tau}{2}.$

The growth of complexity cannot continue indefinitely. Though there is no limit to the size of any individual circuit, if the circuit is too big there will typically be another, shorter, circuit that implements an almost identical unitary. We can estimate how large a circuit can be before it is likely to have been  `short-circuited' by calculating the volume of $\suk$. There are a continuous infinity of elements of $\suk$, so any finite arrangement of discrete gates can only hit measure zero; for this reason, we introduce a tolerance---we settle for getting sufficiently close (in the inner product sense) to the target unitary. An $ \epsilon$-net is a kind of lattice on $\suk$ that divides it into small patches of linear size $\epsilon$. The number of such patches that it takes to cover $\suk$ is double exponential in $K$, of order\footnote{As a function of $\epsilon$, the number of patches is approximately $e^{4^K \log \epsilon}$; in what follows we will ignore the weak dependence on $\epsilon$ and focus on the double exponential dependence on $K$.}
 $e^{4^K}$. Since the number of possible circuits grows exponentially with the depth, this implies that the maximum complexity is exponential in $K$,  
\be 
\CC_{\text{max}} \sim 4^K.
\label{C-max}
\ee

Complexity $\sim 4^K$ is not only the largest possible complexity, it is also the complexity of the overwhelming majority of unitary operators. Thus, while this has not been proved, it is believed that the linear growth in complexity, $K \tau /2$, continues for an exponential time, only appreciably slowing when the complexity gets close to its maximum value. 

 Once the complexity reaches its maximum value it will fluctuate  in the vicinity of $\CC_\textrm{max}$ for a time of order $e^{4^K}$ and then execute quantum recurrences, quasiperiodically returning to small values. The time-scale $\tau\sim e^{4^K}$ is called the quantum recurrence time.

\begin{figure}[H]
\begin{center}
\includegraphics[scale=1.25]{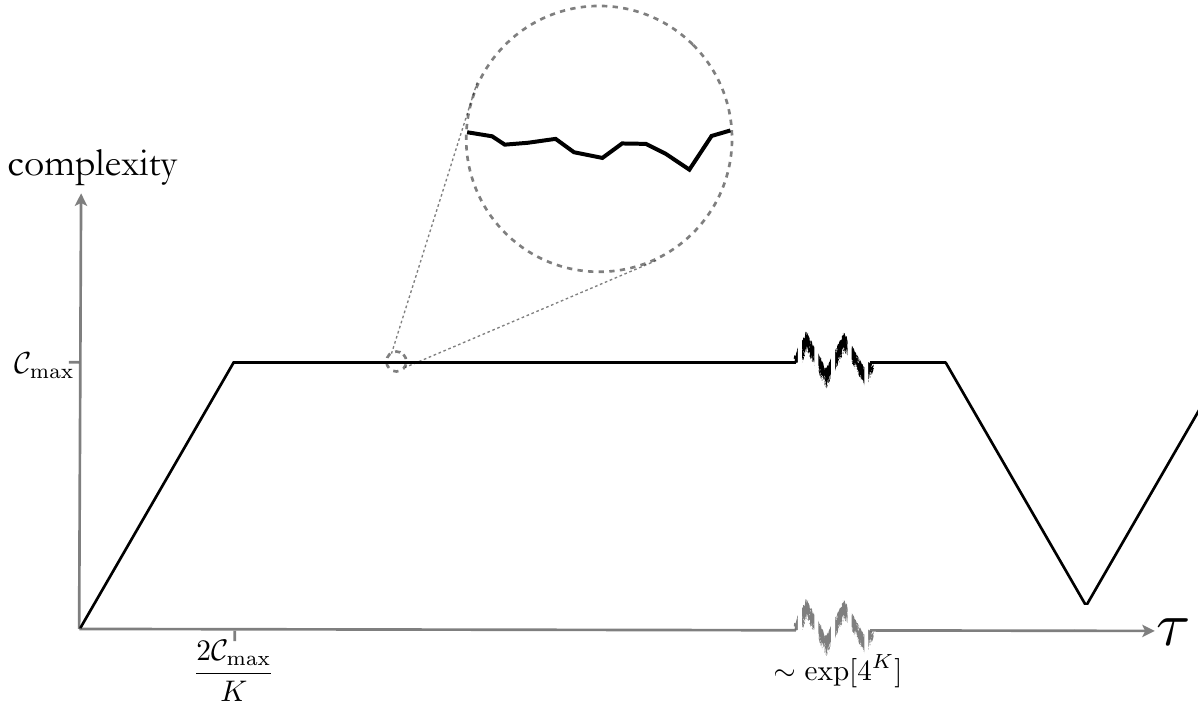}
\caption{The evolution of the computational complexity of the operator $e^{i H \tau}$ for a generic  $k$-local Hamiltonian $H$. At $t=0$ the operator $e^{iHt}$ is the identity  and so has complexity zero. At early and intermediate times the complexity increases linearly, with coefficient $K/2$. After a time exponential in the number of qubits $K$, the complexity saturates at a value $\mathcal{C}_\textrm{max}$ that is exponential in $K$. It then fluctuates near that maximum value. Very very rarely---so rarely that we must wait the double exponentially long quantum recurrence time for it to be likely to have happened even once---the complexity of the system may fluctuate down to near zero, before growing again.}
\label{f1}
\end{center}
\end{figure}

Figure \ref{f1} shows the complexity history, as conjectured in \cite{Susskind:2015toa}, of the evolution operator $U(\tau) = e^{iH\tau}$ for a fast scrambler. 
\bn 

To summarize: the complexity initially increases linearly, with a rate of increase equal to the internal energy \cite{Brown:2015bva,Brown:2015lvg} which we assume is proportional to the number of qubits $K$. The numerical coefficient is dependent on the exact definition of complexity but we will adopt a convention motivated by Eq.~\ref{Ngates}
\be
\CC(\tau) = \frac{K}{2}\tau.
\label{C=Kt/2}
\ee 
Once $\CC(\tau)$ reaches its maximum value a long period of complexity-equilibrium will follow, during which the complexity remains near maximum. On very long time scales, the complexity will fluctuate substantially below its maximum and on double exponentially long time scales very large fluctuations will return $\CC$ all the way back to near zero. Conditional on having backtracked to a low complexity at some time, the system will typically have reached that value via a rate of decrease of  $-K/2.$\\

The growth of complexity has been conjectured to be holographically dual to the growth of the Einstein-Rosen bridge  (ERB)  connecting two entangled black holes \cite{Susskind:2014rva,Stanford:2014jda,Roberts:2014isa}. Evidence for this conjecture is that we see all the same phenomenology for ERBs as we do for complexity. Like complexity, ERBs initially grow linearly with time. Indeed, ERBs continue to grow linearly for as long as classical gravity continues to hold. There are known non-perturbative effects which destroy the validity of classical gravity on a time scale exponential in the entropy \cite{Hawking:1982dh}; if the holographic duality holds up to this time, then linear growth of the ERB continues for an exponential time. 

\subsection{Decreasing Complexity is Unstable} \label{sec:decreasingcomplexity}
We may artificially create a period of decreasing complexity by time-reversing the  quantum circuit, i.e.,   Hermitian conjugating the gates and applying  them in reverse order. This will replace the normal increase of complexity with a reversed history of decreasing complexity.

However, decreasing complexity is unstable: a small perturbation of a single gate (or a thermal photon) will soon cause the complexity to stop decreasing.  After a scrambling time, 
\be 
 \tau_* = \log K,
\ee
the complexity will start increasing, again  with rate $K/2$. 

The random circuit model is useful for seeing why there is a delay time $\tau_*$ before the complexity begins to increase. The scrambling time is defined as the time that it takes for a simple perturbation such as an extra gate to spread through the system affecting every qubit. 
Let us suppose the complexity is large but decreasing, and that at some point we apply an extra single-qubit gate. Initially only one qubit is affected---let us call it ``infected"---by the action of the extra gate. The rest of the system, being uninfected, will continue along its trajectory of decreasing complexity. 
After an additional time-step two qubits will be infected, then four, eight, etc. But for large $K$ this is still a negligible fraction of the system. After  $\tau$ time-steps the epidemic will have spread to 
\be 
s(\tau)=e^{\tau}
\label{size}
\ee
 qubits.\footnote{For $2$-local circuits the size will grow as $2^{\tau}$, where $\tau$ is the number of discrete time-steps. In what follows we'll treat time as continuous and then rescale time to give $e^{\tau}$. In general, we can also consider $k$-local circuits where $k$ is some order one number. By rescaling time and appropriately defining gate complexity,  the complexity of the circuit can always be made to grow as $e^\tau$ at early times; however, the late-time rate at which the epidemic approaches saturation (Eq.~\ref{epi2}) is not robust against changing $k$, and scales instead like the $k$-dependent $e^{-\tau/(k-1)}$. More discussion of the normalization of the time variable $\tau$ and its relation to the various time units used in black hole physics appears in Appendix \ref{appendixaboutunits}. \label{$k$-local}} The notation $s(\tau)$ stands for the \it size \rm of the epidemic after $\tau$ time-steps \cite{Roberts:2014isa}.
 
 This exponential behavior is a sign of chaos and has been studied by Maldacena, Shenker, and Stanford \cite{Maldacena:2015waa}.  They refer to the exponent in the exponential growth formula as the quantum Lyapunov exponent \cite{Kitaev2014talk}.

The scrambling time $\tau_* = \log K$ is the number of steps needed to infect most of the qubits.
Once the scrambling time has passed, the delicate coordination required for the complexity to be decreasing has been completely disrupted by the effects of the perturbation and the normal condition of increasing complexity will have been restored.

By contrast with the decreasing case, increasing complexity is an entirely stable condition.  If the complexity was set to increase, then a perturbation generically won't change that---after a perturbation, the complexity will continue to increase. 

\subsection*{The Second Law of Complexity}

The similarity with the evolution of entropy in classical chaotic systems is obvious and suggests a \it second law of complexity \rm with all the same qualifications as for the second law of thermodynamics. The simplest version would be:

\bn
\it Complexity always increases.\rm

\bn
However, just as it would be for entropy, this is too simple. We can instead try:

\bn
\it Complexity almost always increases.\rm

\bn
But this is also not true: just like entropy, complexity almost always fluctuates about its maximum.

\bn
The correct formulation is:

\bn
\it Conditioning on the complexity being less than maximum, it will most likely increase, both into the future and into the past. 
\rm

\bn
The big difference is that the second law of complexity operates on  vastly longer time scales than its entropic 
counterpart. The time required for the classical entropy to fluctuate down to near zero---the classical recurrence time---is exponential in the entropy $e^K$. By contrast, 
the time required for the quantum complexity to fluctuate to near zero---the quantum recurrence time---is double exponential in the entropy $e^{4^K}$. At the end of Sec.~\ref{sec:relation-nielsen-geometry} we comment on the origin of this similarity between quantum complexity and classical entropy.

\subsubsection{Black Holes and Shock Waves}
In \cite{Hayden:2007cs,Sekino:2008he}\ it was pointed out that the dynamics of a black hole is governed by a $k$-local chaotic Hamiltonian, so we can also study the behavior of complexity in the context of black holes. It was conjectured \cite{Stanford:2014jda,Roberts:2014isa}    that the evolution of complexity is reflected in the growth of black hole interior geometries. By studying the time evolution of Einstein-Rosen bridges, similar behaviors are found as those we expect from quantum circuit complexity. For example, the eternal black hole in AdS has a Penrose diagram which is time-reversal symmetric. During the first half of the evolution the ERB shrinks. This is the white hole era. References \cite{Shenker:2013pqa,Stanford:2014jda,Susskind:2015toa} have studied what happens if during this era a thermal-scale perturbation is applied to the white hole. It is found that the volume (or action) of the ERB continues to decrease for a scrambling time, but then reverses, effectively turning the white hole into a black hole. This exactly parallels the behavior of complexity for chaotic quantum systems. On the other hand a similar perturbation applied during the black hole era has very little effect. In other words, increasing ERB size is a stable condition. 

The most striking evidence for the holographic duality between the complexity of the boundary theory and the size (volume or action) of the corresponding ERB in the bulk is provided by the switchback effect  \cite{Stanford:2014jda,Susskind:2014jwa}. To see this effect, we must study the evolution of precursors.

\subsection{Precursors: Single Perturbations} \label{singleprecursors}
As we just saw, in chaotic systems tiny changes have huge consequences---the passage of time amplifies perturbations until they transform the fate of the whole system. We can  capture this phenomenon by studying `precursors'. A precursor $W(\tau)$ measures the difference between the operator as it is now and what the operator \emph{would} have been by now had we slightly perturbed it at some point in the past. Clearly one way to construct such an operator is to evolve `backwards in time' (i.e. undo the time-evolution that has happened, by acting on it with the inverse of the time-evolution operator), then hit it with the small perturbation, and then run it back to the present by acting with the normal forward-directed time-evolution operator:
\be 
W(\tau) = U(\tau) W U^{\dag}(\tau).
\ee
Here $U$ is the time evolution operator $U(\tau) = \exp(-iH\tau)$ and $W$ is a simple operator of unit complexity. (For example, for a quantum circuit $W$ could be a single-qubit Pauli operator.) 

Let us now consider the complexity of $W(\tau)$. For $\tau = 0$ the complexity is tiny
\begin{equation}
\CC[W(\tau=0)] = \CC[W] = 1.
\end{equation}
However, the complexity of $W(\tau)$  grows with $|\tau|$, as the effect of the initially-small perturbation cascades through the system. An upper bound on the growth rate is given by the triangle inequality
\begin{equation}
\CC[W(\tau)] \ \ \leq  \  \  \CC[U(\tau)] + \CC[W] + \CC[U^\dagger(\tau)] \ = \  \frac{1}{2} K |\tau| + 1 + \frac{1}{2} K |\tau| \ \  \sim  \  \ K  |\tau|, \label{naive}
\end{equation}
since the minimal circuit that implements $W(\tau)$ can certainly be no bigger than the circuit formed by  concatenating the individual circuits for $U$, $W$ and $U^\dagger$.

This naive concatenated circuit successfully makes $W(\tau)$, but it is not the smallest circuit that makes $W(\tau)$. This is most easily seen considering the case where $W$ is the simplest of all operators, namely the identity operator. In that case $U$ and $U^{\dag}$ cancel, so $W(\tau)$ is also the identity operator and has complexity zero. 

The point is similar to the earlier discussion of the instability of decreasing complexity. The operator $W$ affects only a single qubit; in other words it acts as the identity on all but one qubit. Therefore there is a large amount of cancellation
between the gates of $U$ and $U^{\dag}$ until the effect of $W$ has spread through the system. This partial cancellation leads to the minimal circuit being smaller than the naive concatenated circuit by an amount $K \tau_*$, a phenomenon known as the `switchback effect'. The switchback effect is one of the signature phenomena of the evolution of quantum complexity, and reproducing it will be a strong test of our analog model in Sec.~\ref{sec3}.

\vspace{1.3cm}

To be more quantitative about how the complexity of $W(\tau)$ changes with time, we can consider the random circuit model. Going back to Eq.~\ref{size} for the size of the epidemic, it is obvious that the size cannot grow for too long since the epidemic must saturate once every qubit is infected. Taking into account that only uninfected qubits can become infected, one finds that for large $K$  the size of the epidemic satisfies the differential equation  \cite{Susskind:2014jwa}
\be 
\frac{ds(\tau)}{d\tau} = s\frac{K-s}{K-1}.
\label{epidemic}
\ee
The solution  is the `epidemic' function 
\be  
\ \ \ \ \ \ \ \ \ \ \  \ \ \ \ \ \ \ \ \ \ \ \ \ \ \ \ \ \ \ \ \ \ \ \ \ \ \ \ \ \ \ \ \ s(\tau) = \frac{Ke^{\tau}}{K+ e^{\tau}}  = K\frac{e^{(\tau-\tau_*)}}{1+e^{(\tau-\tau_*)}} \ \ \ \ \ \ \ \ \ \ \ \ \ (\textrm{with } \tau_* = \log{K}).
\label{epi2}
\ee
Several things are seen from these equations. The early exponential growth of Eq.~\ref{size} is recovered for time  less than $\log{K}.$ At later time, every qubit is infected, and so the size of the epidemic saturates\footnote{The saturation of the size of the epidemic is not connected with the saturation of complexity  at its maximum value in Eq.~\ref{C-max}. The saturation of size occurs at the scrambling time, which is logarithmic in $K$; the saturation of complexity occurs at a time exponential in $K$.} at $K$. Equation \ref{epi2} shows saturation takes a scrambling time. This formula was more rigorously justified by a model using a random time-dependent Hamiltonian in \cite{Shenker:2014cwa}.

\begin{figure}[H]
\begin{center}
\includegraphics[scale=.55]{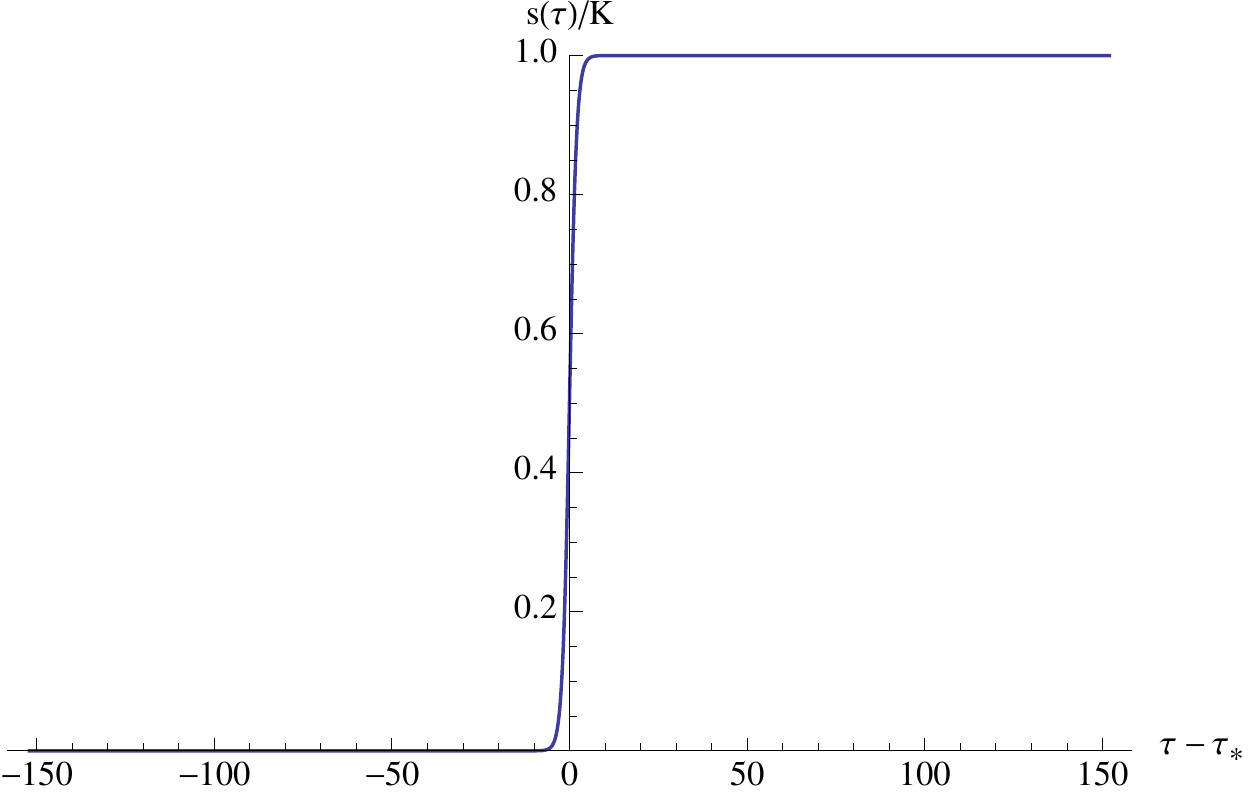}
\caption{The size of the precursor as a function of $\tau-\tau_*$ for the epidemic function Eq.~\ref{epi2}. For $\tau - \tau_* \ll 1$ the size is growing exponentially as the epidemic spreads throughout the system, but it's growing from such a small baseline as to be visually indistinguishable from zero. When $\tau = \tau_*$ almost every site is infected and $s(\tau)$ abruptly saturates at $K$.}
\label{sizeepidemic}
\end{center}
\end{figure}

Figure \ref{sizeepidemic} plots of the size of the precursor as a function of time, $s(\tau)$. At early times the precursor is growing exponentially, but is still too tiny to be visibly distinct from zero. After a scrambling time the size rapidly grows from near zero to near one, saturating in a few thermal times. (See Appendix \ref{appendixaboutunits}.) When there is a large hierarchy between the scrambling time and the thermal time, which is to say when $\log K \gg 1$,  the transition is quite sharp and almost looks like a step function. 

From $s(\tau)$ it is straightforward to calculate $\CC(\tau)$. We call those gates acting on infected qubits `infected gates'. Recall from the discussion below Eq.~\ref{naive} that the complexity of a precursor  is the number of uncanceled  gates. It can also be identified with the sum over time of the number of infected gates at that time. On the other hand, at each time-step the number of infected gates is equal to the number of infected qubits,\footnote{This statement is true for $2$-local circuits. For general $k$-local circuits we need to rescale time. See earlier footnote \ref{$k$-local}.} which is the size of the precursor. It follows that the complexity is the integral over time of the size. Or more simply,
\be 
 \frac{d\CC(\tau)}{d\tau} = s(\tau). \label{eq:precursorsizedefinition}
\ee
Combining this equation with Eq.~\ref{epi2} gives
\begin{align}
\CC(\tau)=K\log\left(1+e^{\tau-\tau_*}\right)=\begin{cases}
e^{\tau} & \textrm{for } \tau_*  - \tau \gg 1\\
\frac{K}{2}(2\tau-2\tau_*) & \textrm{for } \tau - \tau_*  \gg 1 \  .
\end{cases}
\label{switchback}
\end{align}

At early time the complexity increases exponentially with Lyapunov exponent $1$. As explained in the appendix, this is the value for the quantum Lyapunov exponent expected from the arguments of \cite{Maldacena:2015waa}.
After a scrambling time the complexity grows linearly. The delayed onset of linear growth causes the complexity to be $K \tau_*$ less than it would have been with no delay.

\subsubsection{Single Precursors and Black Holes}
We have seen how precursor perturbations effect the growth of quantum complexity. The growth of quantum complexity is believed to be holographically dual to the growth of wormholes behind a black hole horizon.
For a two-sided AdS black hole, the dual of applying a precursor $W(t_w)$ to the boundary state is throwing in a thermal graviton at time $-t_w$. At first the photon has little effect---its energy is tiny. However, as it falls into the AdS gravitational potential, it blue-shifts exponentially. Soon, the little graviton has grown to a mighty shockwave, which  greatly distorts the geometry of the black hole interior. A classical gravity calculation \cite{Shenker:2013pqa} calculates the size of the backreaction caused by early-time shockwaves as\footnote{To get Eq.~\ref{sizegravity} from \cite{Shenker:2013pqa} we identify complexity with the geodesic length of wormhole. It can be shown  \cite{toappear} that Eq.~\ref{sizegravity} will hold true regardless of which detailed prescription of bulk dual of complexity we use. In particular, this functional form will be true if we use ERB volume \cite{Susskind:2014rva,Roberts:2014isa} or 
action \cite{Brown:2015bva,Brown:2015lvg}.} 
\begin{equation}
s[W(t_w)]=\frac{d\CC[W(t_w)]}{dt_w}=K \frac{c \, e^{\frac{2\pi}{\beta}(t_w-t_*)}}{1+ c \,  e^{\frac{2\pi}{\beta}(t_w-t_*)}},
\label{sizegravity}
\end{equation}
where $c$ is an order $1$ constant that is proportional to the exact energy of the disturbance in units of temperature. This has exactly the same functional form as Eq.~\ref{epi2}.\footnote{Note the prefactor $\frac{2\pi}{\beta}$ in front of the time $t_w$. The $t_w$ is measured in Schwarzschild time, and the $\frac{2\pi}{\beta}$ transforms it into Rindler time, which corresponds to the dimensionless  $\tau$  used in the circuit model.} 
This is an important consistency-check for the holographic complexity-geometry correspondence, and we see that the correspondence passes: the picture of epidemic spreading successfully captures the scrambling properties of a black hole horizon.

\subsection{Precursors: Multiple Perturbations} \label{subsec:multipleprecursors}

\begin{figure}[H]
\begin{center}
\includegraphics[scale=.65]{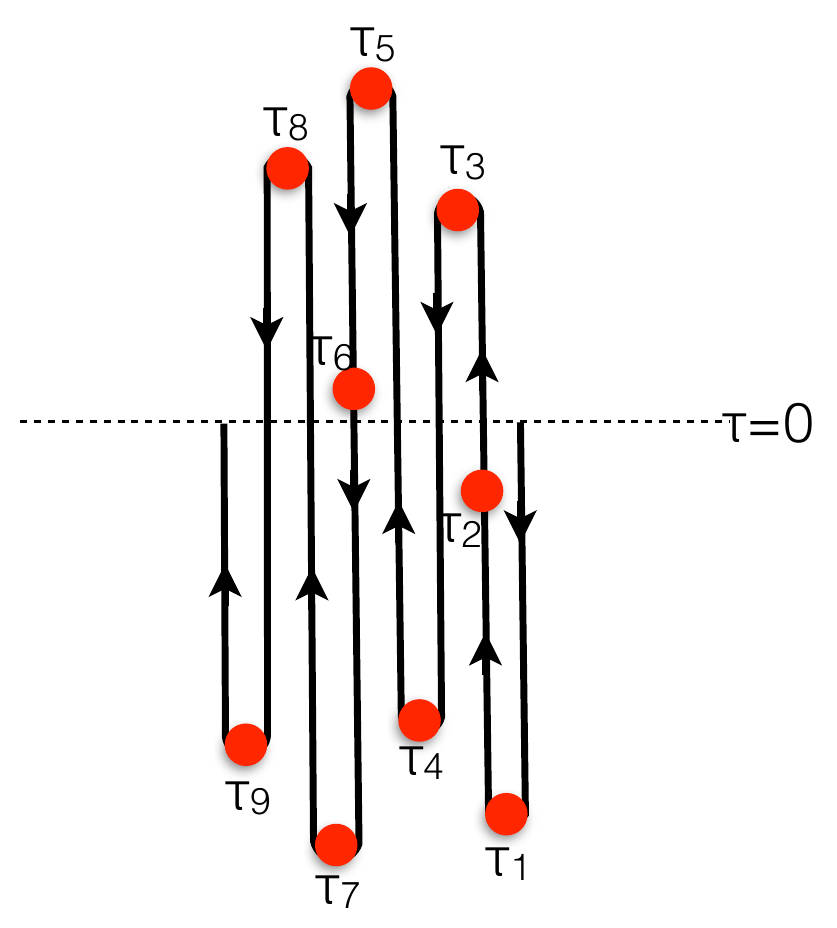}
\caption{The multiple precursor operator $W_\textrm{multi}(\tau_n, \tau_{n-1}, \ldots ,  \tau_1)$. When $\tau_i - \tau_{i-1}$ and $\tau_{i+1} - \tau_i$ have the same sign, as at $\tau_2$ and $\tau_6$, $W_i$ is called a `through-going' insertion and there is generically no cancellation. When $\tau_i - \tau_{i-1}$ and $\tau_{i+1} - \tau_i$ have opposite signs, as at the other $\tau_i$, this is called a `switchback' and there is a partial cancellation that reduces the complexity by $K \tau_*$.}
\label{multiprecursor}
\end{center}
\end{figure}

Rather than perturbing the system once, we may perturb the system multiple times. We can implement such a perturbation with \cite{Shenker:2013yza}
\begin{eqnarray}
W_\textrm{multi}(\tau_n, \tau_{n-1}, \ldots ,  \tau_1) & \equiv &W_n(\tau_n)W_{n-1}(\tau_{n-1})...W_2(\tau_2)W_1(\tau_1)  \\
& = &e^{-iH\tau_n}W_ne^{iH(\tau_n-\tau_{n-1})}W_{n-1}e^{iH(\tau_{n-1}-\tau_{n-2})}...W_2e^{iH(\tau_2-\tau_1)}W_1e^{iH\tau_1},\nonumber
\end{eqnarray}
where the $\tau_i$'s may not be chronologically ordered. Pictorially, we represent such a product by a time-fold:  see Fig.~\ref{multiprecursor}, where each red dot represents a small perturbation $W_i$, and the arrows indicate the order in which the perturbations act.
When the perturbations are well separated, so that the influence of $W_i$ has  spread throughout the whole system before $W_{i+1}$ acts, we expect one switchback-subtraction at each time reversal. This means the complexity of this operator will be
\begin{align}
\label{multiswitchback}
\CC[W_\textrm{multi}(\tau_n, \tau_{n-1}, \ldots ,  \tau_1)]=\frac{K}{2}(\tau_f-2n_{sb}\tau_*),
\end{align}
where $n_{sb}$ is the number of switchbacks and $\tau_f$ is the total `folded time'
\begin{equation}
\tau_f  \equiv |\tau_1 - \tau_2| + |\tau_2 - \tau_3| + \ldots + |\tau_{n-1} - \tau_n|.
\end{equation}

The number of switchbacks $n_{sb}$ is generally not the same as the number of perturbations. To get a switchback subtraction, the Hamiltonian evolution before and after the perturbation must go in opposite directions; it is only then that there is any cancellation. When the perturbation is a through-going insertion, such as at $\tau_2$ and $\tau_6$ in Fig.~\ref{multiprecursor}, there is no cancellation. Through-going perturbations still completely change the microstate (after long enough time), but they have little effect on the time evolution of complexity.

\subsubsection{Multiple Precursors and Black Holes}
Products of precursors  give stringent tests of the holographic duality between quantum complexity and the classical geometry of ERBs. In the black hole context one studies this by studying black hole geometries perturbed by multiple shockwaves separated by large times \cite{Shenker:2013yza}; sure enough the complexity results Eq.~\ref{multiswitchback} and  gravity results \cite{Stanford:2014jda} agree.

\sc
\section{Analog Model: Particle in Hyperbolic Space} \label{sec3}

 In the last section, we considered the evolution of the complexity of a quantum system. In this section we will consider an analog model---a classical particle moving on a two-dimensional negatively curved space of large genus. We will find a surprisingly detailed
parallel between the two.

\subsection*{The Analog Model}

\begin{figure}[H]
\begin{center}
\includegraphics[scale=.2]{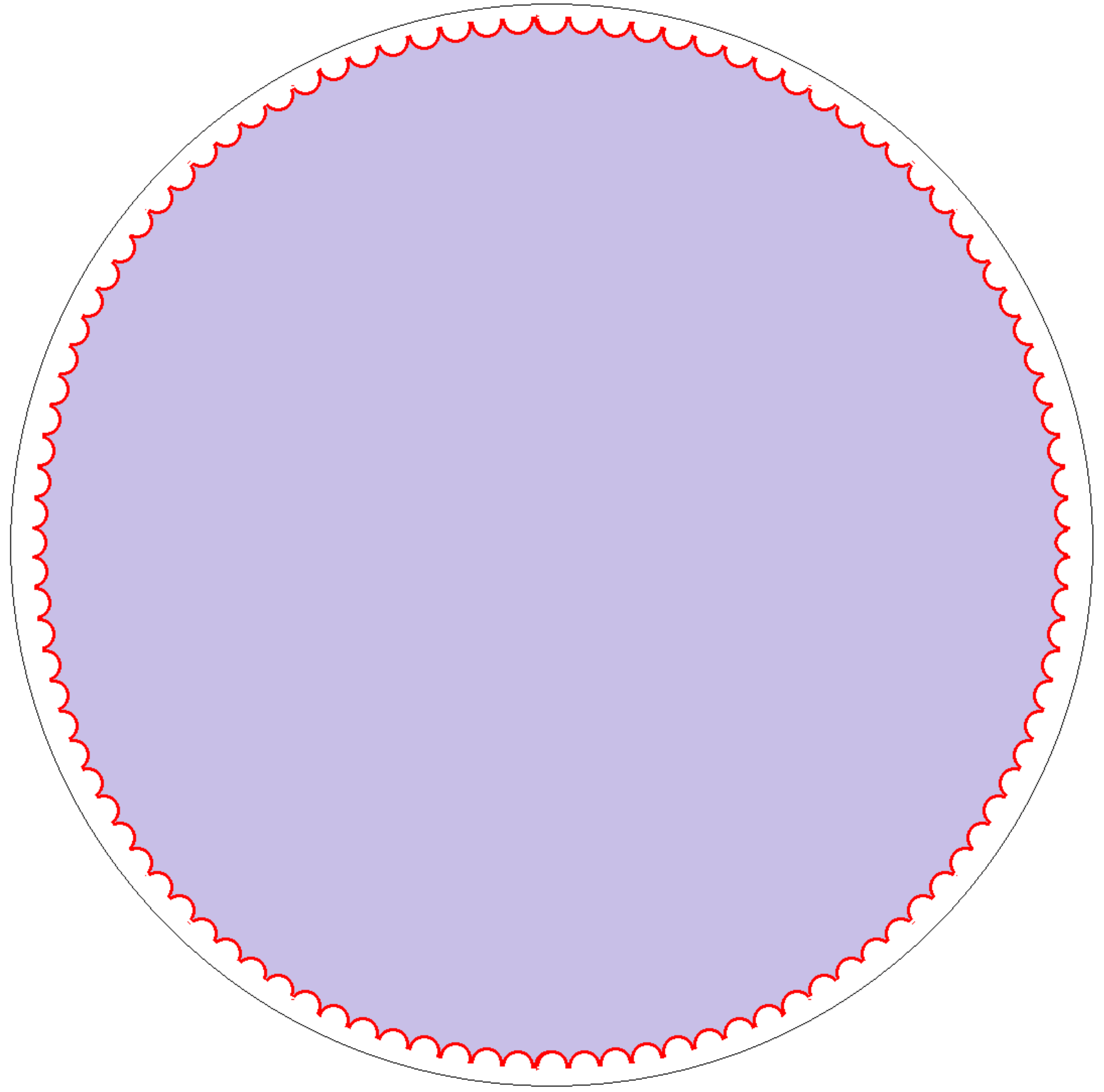}
\caption{An example of a compactification of the hyperbolic plane. Start with a ginormous equilateral polygon centered at the origin, and then identify sides to make the compact and conical-deficit-free space $\CH_g$.}
\label{f2}
\end{center}
\end{figure}

The analog model features a non-relativistic particle moving on a uniformly negatively curved geometry. An infinite hyperbolic plane $\mathbb{H}^2$ with curvature length $K/2$ has metric  
\begin{equation}
dl^2 = \frac{K^2}{4}  (dr^2 + \sinh^2{r} d\theta^2 ) . \label{metric-r-theta}
\end{equation}
The Gaussian curvature is $-4/K^2$. The origin, at $r=0$, will correspond to the identity operator in the qubit system of the last section. The volume $V$ within a distance $L$ of the origin is (for $L \gg K$)
\begin{equation}
V \sim K^2 e^{2L/K}. \label{eq:Vexp}
\end{equation}
We will compactify the hyperbolic plane  to a uniformly negatively curved space of genus $g.$ We denote the compactified space $\CH_g$. The Gauss-Bonnet theorem implies the genus and the volume are approximately the same. 
In the analog model, 
they will both be double-exponentially large
\be 
V \sim g \sim e^{4^K}. \label{eq:vsimg}
\ee
%
%
%
%
%
To implement the compactification, we start with an equilateral hyperbolic polygon with $4g$ sides centered at the origin. This is shown in Fig.~\ref{f2}. We then pair up the sides of the polygon, and identify the elements of each pair. The distance of the polygon from the origin is constrained by the condition that the outer points fit together without a conical singularity; given Eqs.~\ref{eq:Vexp} and \ref{eq:vsimg} this distance must be approximately $4^K.$

%
\vspace{-5mm}

The analog model that we will present is the motion of non-relativistic particles on $\CH_g$.
Unperturbed particles follow geodesics of $\CH_g$ with uniform velocity. (In the Poincar\'e disk, these geodesics are circles centered on the boundary.) When impulsively perturbed, the particles are deflected. As we will see, both  perturbed and unperturbed motion in our analog model closely resembles the evolution of complexity of quantum system.  The resemblance involves the identifications:
 \begin{center}
\begin{tabular}{c||c} 
COMPLEXITY \  \  & \ \    HYPERBOLIC SPACE  \Bo \\
\hline
\hline
number of qubits $K$ & curvature length ${K}/{2}$\To \Bo \\
\hline
evolution generated by  a 2-local & geodesic motion on $\CH_g$ \To  \\ 
time-independent Hamiltonian $H$ &   \Bo  \\
 \hline
identity operator  & center of the Poincar\'e disk  \To \Bo \\
\hline
trajectory in $\suk$ swept out by &  particle trajectory through $\CH_g$  \To \\
the time evolution operator $e^{-i H t}$ &  along a geodesic with velocity $K/2$   \Bo  \\
\hline
complexity of the operator $U( t)$ & distance of the particle from origin \To \\ 
(number of gates in minimal circuit) & (length along minimal geodesic)   \Bo  \\
\hline
maximum complexity  & length, $L$, of the longest minimal   \To \\
(this is of order $4^K$, see Eq.~\ref{C-max})& geodesic from the origin  \Bo  \\
\hline
number of unitary operators in $\suk$ & volume, $V$, of $\CH_g$ \To \\
 (this is of order $e^{4^K}$)&  (this is of order $e^L$)   \Bo \\
\hline
 perturbation with  & a perpendicular small displacement   \To \\
simple operator $W$ &  of trajectory, as in Fig.~\ref{kick}    \Bo  \\
\hline
decreasing complexity  & decreasing distance from   \To \\
unstable  & origin unstable \Bo \\
\hline
quantum Lyapunov exponent  & classical Lyapunov exponent    \To \\
of fast scrambler & of motion on $\mathbb{H}^2$ \Bo \\
\end{tabular}
\end{center}

\newpage

\subsection*{Recurrences}

\begin{figure}[H]
\begin{center}
\includegraphics[scale=.24]{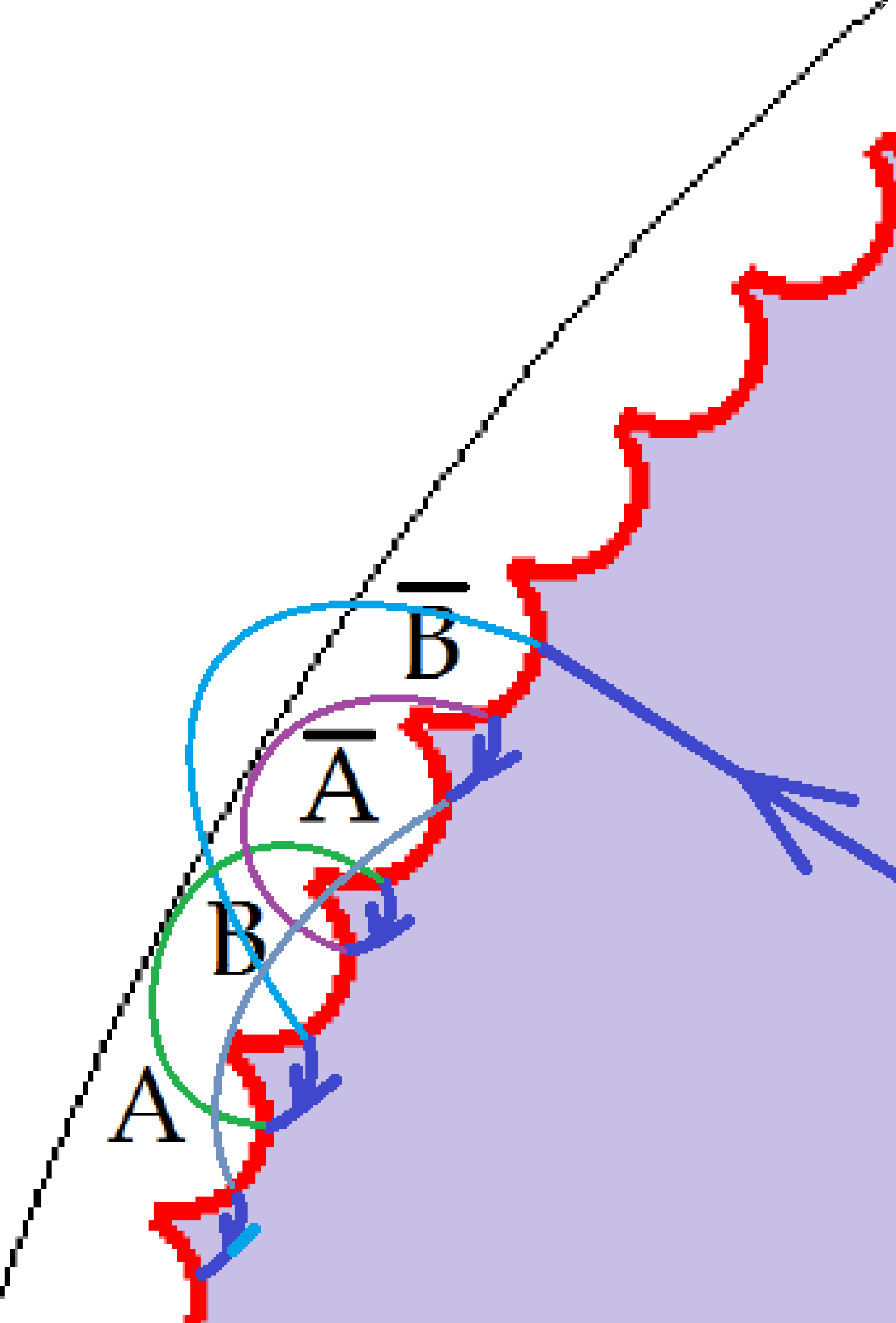}
\caption{Following a geodesic on $\CH_g$. When the geodesic crosses the red polygon that bounds $\CH_g$ through side $A$, it re-enters the geometry at the identified point on the polygon through side $\bar{A}$. (For convenience, the identified sides are shown here as being close together, though typically they will be very distant.) In this picture, a geodesic is traced through several such exits, re-entries, and re-exits.}
\label{f5}
\end{center}
\end{figure}

When an outgoing trajectory of increasing radius (increasing complexity) reaches a side of the polygon it jumps to  the identified side and reenters the geometry. Immediately after re-entry it is traveling towards smaller $r$, implying that the complexity is temporarily  decreasing. Generically the decrease lasts for a short time and then turns around so that the trajectory soon hits another nearby side. This leads to a fluctuating complexity that tends to fluctuate near the maximum value for a very long time. Figure \ref{f5} shows a portion of the geodesic after it first passes through one of the polygon sides.

After re-entering the shaded region the geodesic will  soon exit again. Typically it will re-exit at a  neighboring side. Once in a while the angle will be such that a side or two will be skipped over. It takes an extreme fine tuning for the geodesic to return to anywhere near the origin, i.e., to low complexity.

It is easy to calculate the fine-tuning of the angle of re-entrance required for the complexity to reach a sub-exponential  value. Relative to the radial direction the angle must be fine-tuned to 
\be 
\delta \theta \sim \exp\left(-4^K\right).
\ee
Indeed, we can calculate the probability for this motion to reach a  distance from the origin $\Delta \CC$ less than the maximum. Motion on $\CH_g$ is known to be ergodic and fills the space uniformly, so the probability of reaching a given small radius is inversely proportional to the volume contained within that radius. Using Eq.~\ref{eq:Vexp} this gives
\begin{equation}
\textrm{probability of reaching radius } L - \Delta \CC \ \ = \ \ e^{- 2 \Delta \CC/K}. \label{eq:submaximalcomplexity}
\end{equation}
The expected time for a recurrence is of order the quantum recurrence time of the circuit model, $e^{4^K}$. Thus the model reproduces the expected complexity evolution, Fig.~\ref{f1}.

\subsection{Decreasing Radius is Unstable}

In Sec.~\ref{sec:decreasingcomplexity} we saw that fine-tuned evolutions in which the complexity is decreasing are unstable. Any small perturbation---even that due to a single gate---will generically cause the complexity to stop decreasing and start increasing again. We also saw that the complexity increase is a delayed reaction that sets in about a scrambling time after the perturbation. In this subsection we will see that the same is true in the analog model. We will see that geodesics of decreasing radius (decreasing distance from the origin) are unstable. Any small deflection to the trajectory grows due to geodesic deviation, and after a distance $\frac{1}{2} K \log K$ the trajectory will typically be heading away from the origin again.

\begin{figure}[H]
\begin{center}
\includegraphics[scale=.18]{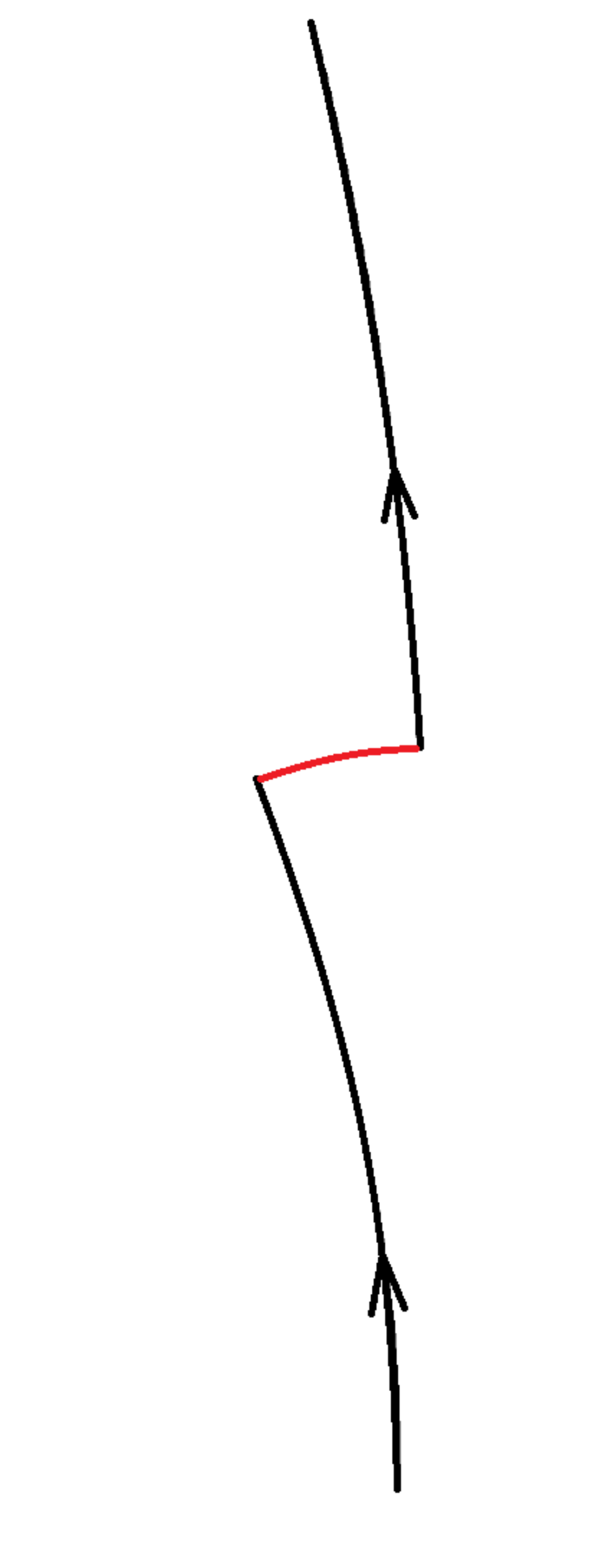}
\caption{A small perturbation represented as a perpendicular kick to a geodesic.}
\label{kick}
\end{center}
\end{figure}

In what follows it will be more convenient to represent $\mathbb{H}^2$ by the Poincar\'e half-plane rather than the disc.
In these coordinates, the metric takes the form,
\be  
dl^2 = \frac{K^2}{4}\left( \frac{dx^2 + dy^2}{y^2} \right).
\ee
 All geodesics are  circles centered on the boundary at $y=0$ (when the circles are infinitely big this gives straight lines in the $y$ direction). It is convenient to map the origin of the Poincar\'e disk, $r=0$, to very large $y$. In the region shown in Figs.~\ref{kick} and \ref{f3}, large $y$ corresponds to small complexity and small $y$ corresponds to large complexity. 

The undisturbed evolution of a quantum system by a time-independent 2-local Hamiltonian  corresponds to geodesic motion on $\mathbb{H}^2$. We will also want to consider disturbances---for the circuit model this might be the action of an extra gate, for the black hole this might be the addition of a thermal-scale perturbation. In our analog model, disturbances will correspond to shunts that displace the particle. For example, the analog of a perturbation by an orthogonal perturbing Hamiltonian\footnote{The inner product is defined by trace norm: $\langle H_1,H_2 \rangle=\text{tr}(H_1H_2)$. On the subspace of simple operators (2-local) this is the same as in Nielsen's geometry \cite{2007quant.ph..1004D}. With this norm, we expect the inner product between a generic $2$-local Hamiltonian and a simple operator to be of order $\frac{1}{K}$ when both are normalized.\label{innerproduct}} 
is two  right-angled turns in quick succession---it first  kicks the particle onto a perpendicular geodesic for a small distance of order unity (corresponding to a complexity of a single gate) before then kicking it again. This is shown in Fig.~\ref{kick}.
Let's use this model of perturbations to recover the result of Sec.~\ref{sec:decreasingcomplexity}---that increasing complexity is stable, but decreasing complexity is unstable.

\begin{figure}[H]
\begin{center}
\includegraphics[scale=.7]{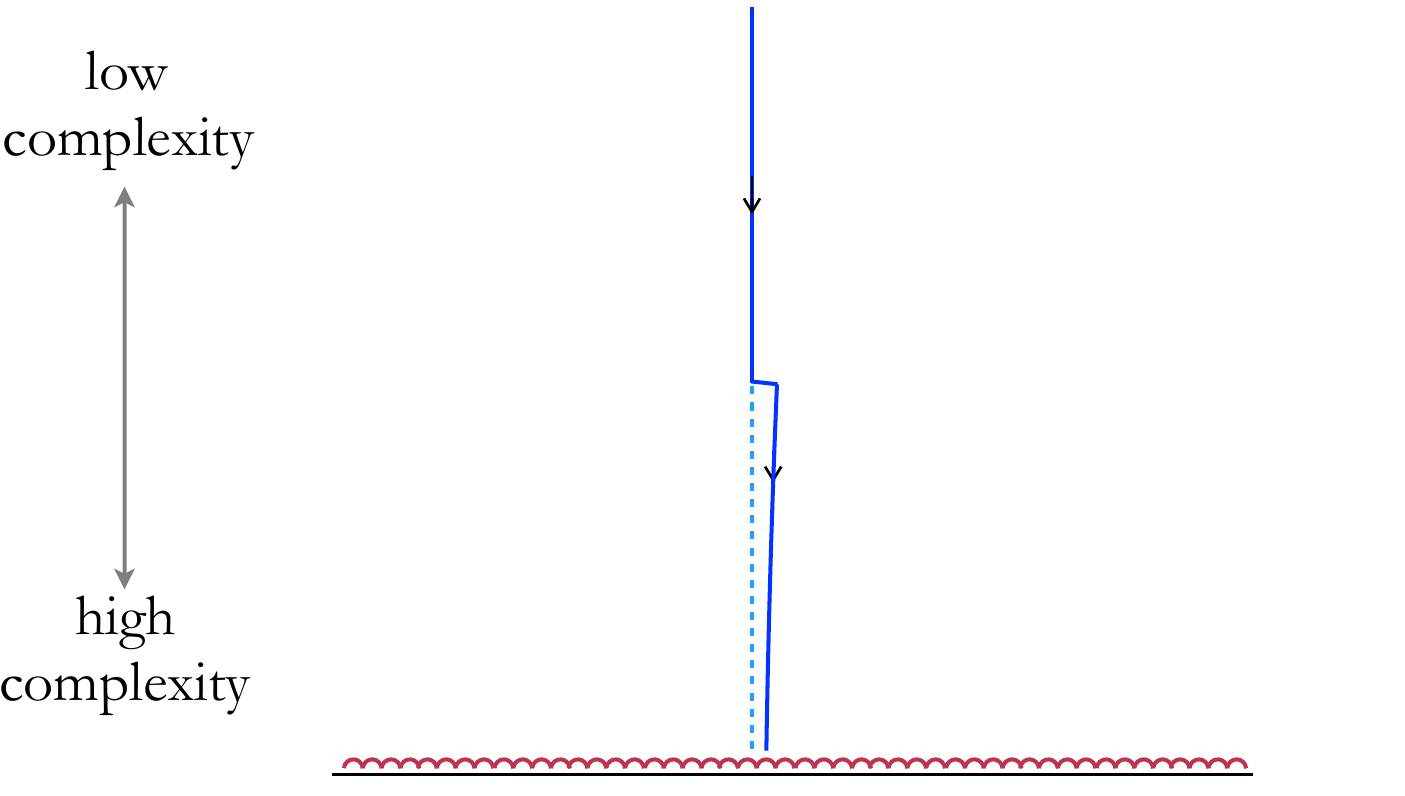}
\caption{Perturbing a trajectory of increasing complexity results in a new trajectory on which the complexity is still increasing. Since the conformal factor $y^{-2}$ is blowing up at small $y$, the horizontal deviation is exponentially increasing. However, the rate of change of complexity is almost unaffected by the kick.}
\label{f4}
\end{center}
\end{figure}

\begin{figure}[H]
\begin{center}
\includegraphics[scale=.22]{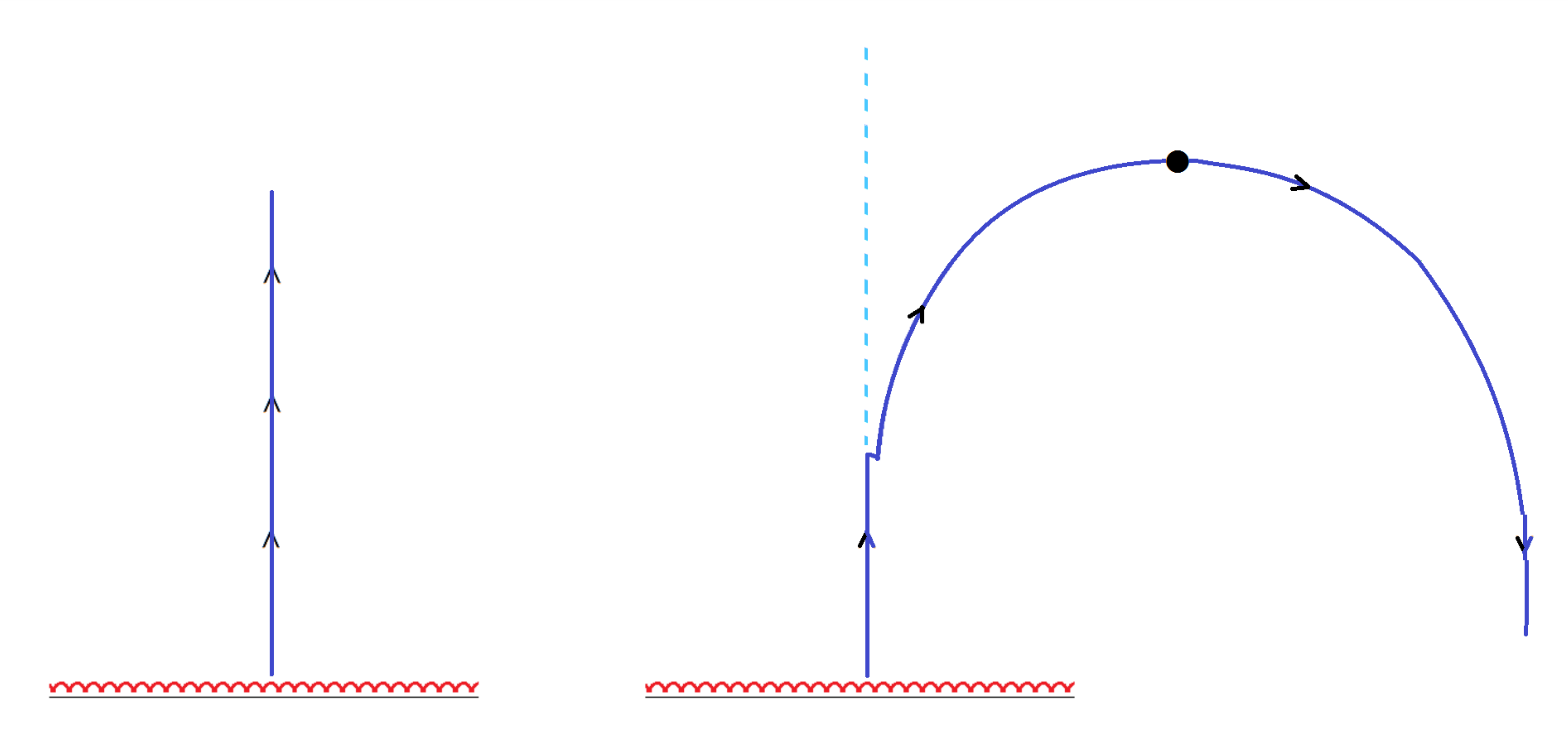} \ \ \ 
\caption{On the left side, we show a trajectory of decreasing complexity---$y$ is increasing. On the right side, we show the effect of a perturbation. The kick results in a small change of direction in the trajectory, which is then magnified by the negative curvature. After a scrambling time, the complexity bottoms out and starts to increase ($y$ decreases). }
\label{f3}
\end{center}
\end{figure}

Figure \ref{f4} illustrates that increasing complexity is stable. A downwards pointing trajectory corresponds to increasing complexity. Consider a kick produced by an extra gate of unit complexity. The kick travels for a unit distance along a geodesic orthogonal to the initial direction, which in the Poincar\'e plane is a small circular arc. (See Fig.~\ref{f3}). Because the kick is represented by a circular arc, after the perturbation the particle is no longer traveling exactly vertically. Thus perturbing this trajectory with a horizontal displacement gives rise, eventually, to a large horizontal metric displacement. However, the new trajectory has almost the same vertical progression as the old trajectory---the complexity was increasing before the kick, and it continues to increase with almost the same rate after the kick. 
(This is an example of the `through-going' insertion discussed earlier.) This behavior thus matches the expectations for circuits and black holes.

Now consider a fine-tuned situation in which the complexity starts large and decreases toward zero. In the analog model, this corresponds to motion in which a particle starts low and proceeds vertically upwards. Once again we may perturb this with a small kick, and once again after the perturbation the particle is no longer traveling exactly vertically. Due to the negative curvature the deviation from vertical grows exponentially.  Eventually the trajectory reaches a maximum height (corresponding to a minimum complexity, at the black dot)  and then falls back in the direction of decreasing $y$ (and hence increasing complexity).

At the turn-around point---the top of the trajectory---the complexity reaches a minimum. The length of the rising portion of the circular arc is easily calculated as
\be 
\Delta l = \frac{1}{2} K \log{K}.
\ee
This corresponds to a time equal to the scrambling time $\tau_*=\log{K}.$ In other words, at the scrambling time the complexity stops decreasing, and begins to increase.  This agrees perfectly with expectations from circuits and black holes.

We find it interesting and suggestive that the so-called quantum Lyapunov exponent \cite{Maldacena:2015waa}, which is by no means a Lyapunov exponent of the classical gauge theory, is a classical Lyapunov exponent of the analog model. 

\bn

\subsection{Precursors: Single Perturbations}

\begin{figure}[H]
\begin{center}
\includegraphics[scale=.4]{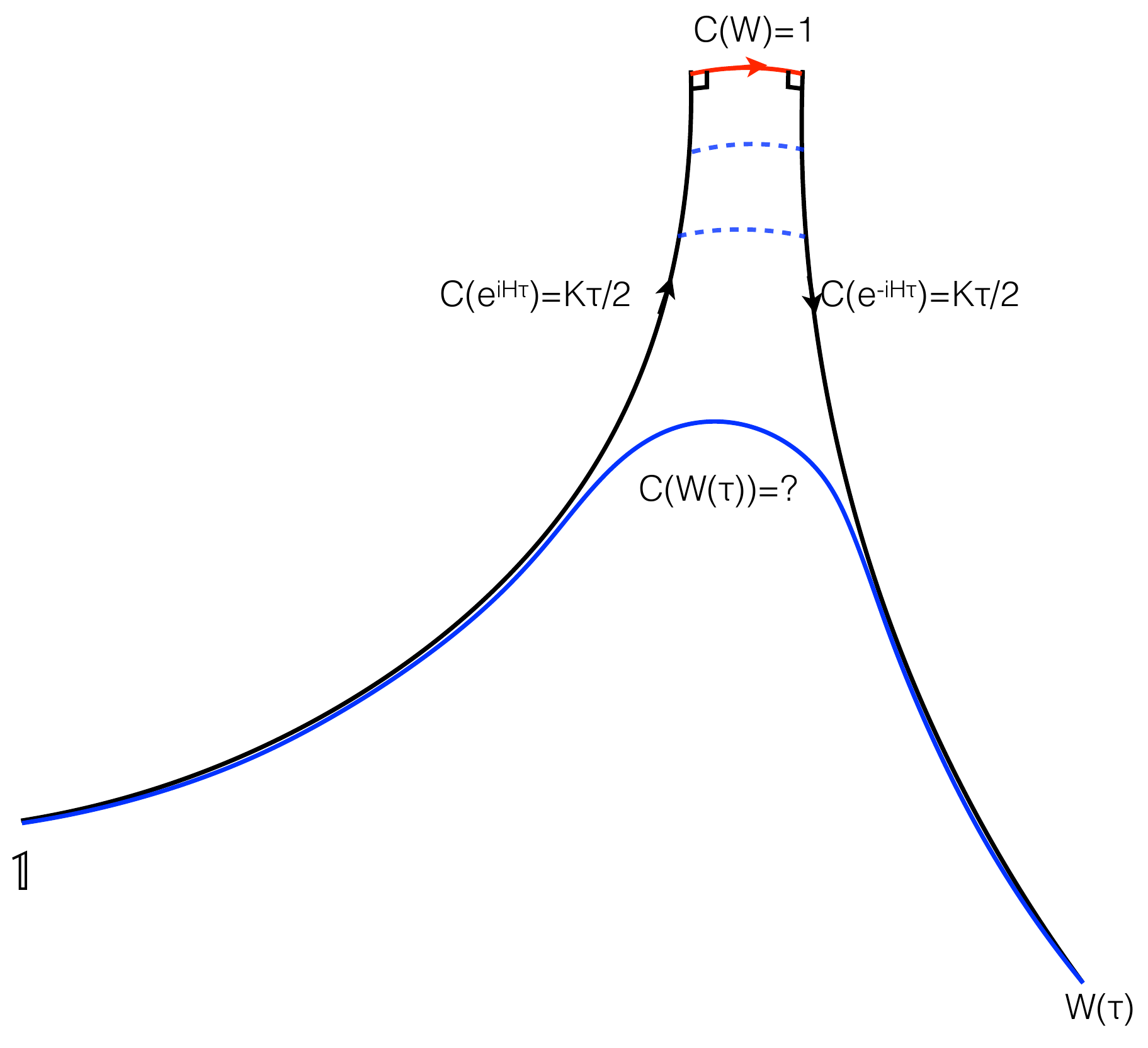}
\caption{A precursor is represented by three segments of geodesics in $\mathbb{H}^2$. This is a schematic picture that represents distances more faithfully than in Poincar\'e coordinates.}
\label{oneprecursorsurface}
\end{center}
\end{figure}

Next, we study precursors in the analog model. The precursor $W(\tau)=e^{-iH\tau}We^{iH\tau}$ is represented by three geodesic segments representing $e^{iH\tau}$, $W$ and $e^{-iH\tau}$; the geodesics representing $e^{\pm iH\tau}$ are in black, and the small segment representing $W$ is in red, as in Fig.~\ref{oneprecursorsurface}. We have seen that the analog of complexity is geodesic length, so the analog of the complexity of $W(\tau)$ is the length of the shortest path from the origin to the point representing $W(\tau)$.

An upper bound on the length of the shortest path connecting the endpoints is given by the triangle inequality
\begin{equation}
\CC[W(\tau)] \ \ \leq  \  \  \CC[U(\tau)] + \CC[W] + \CC[U^\dagger(\tau)] \ = \  \frac{1}{2} K |\tau| + 1 + \frac{1}{2} K |\tau| \  \sim    \ K  |\tau|, \label{alsonaive}
\end{equation}
since the minimal path can certainly be no longer than the path formed by concatenating the individual lines for $U$, $W$ and $U^\dagger$. This is analogous to the naive upper bound on the complexity of the operator $W(\tau)$ from concatenating sub-circuits, as  given in Eq.~\ref{naive}.

This naive concatenated line is not the shortest line that connects the endpoints. Instead, the  minimal geodesic takes a shortcut at the corner. We will see that the length shaved by this shortcut precisely reproduces the switchback subtraction from Sec.~\ref{singleprecursors}.

In Appendix~\ref{appendixhyperbolicspace}, the geodesic distance between the point representing the identity and the point representing $e^{-iH\tau}We^{iH\tau}$ is calculated to be
\begin{equation}
\label{oneprecursor}
\cosh\frac{2\CC[W(\tau)]}{K}=\cosh^2\Bigl[ \frac{1}{K} \Bigl] + \sinh^2\Bigl[ \frac{1}{K} \Bigl] \cosh \Bigl[2\tau \Bigl] .
\end{equation}
From this we can extract both the early time and the late time behavior
\begin{align}
\CC[W(\tau)]=\begin{cases}
\frac{1}{2}  e^{\tau} & \textrm{for } \tau_*  - \tau \gg 1\\
\frac{K}{2}(2\tau-2\tau_*) & \textrm{for } \tau - \tau_*  \gg 1 \  .
\end{cases}
\end{align}
We can also calculate the `size' of the precursor, defined as the rate of change of complexity. For $\log K \gg 1$ this is 
\begin{align}
{s[W(\tau)]} \equiv \frac{dC[W(\tau)]}{d\tau}=\frac{K \sinh \tau}{\sqrt{K^2+\cosh^2 \tau}}. \label{eq:sizeofprecursor}
\end{align}
When $1<\tau<\tau_*$ the size of the precursor increases exponentially with exponent one; after a scrambling time $\tau>\tau_*$ it approaches saturation as $K(1-2K^2 e^{-2\tau}+ \ldots)$.

We see that the evolution of the length of a precursor geodesic is almost-but-not-exactly identical to the evolution of the complexity of a precursor operator in Eq.~\ref{switchback}. For $\tau < \tau_*$ they both grow exponentially with exponent one. At $\tau = \tau_*$ both jump rapidly from near zero to near saturation. And for $\tau > \tau_*$ both grow linearly with the same coefficient, and with the same constant subtraction---what for the operator was a shortening of the naive circuit due to the partial cancellation of the gates is for the geodesic a shortening of the naive length due to taking a shortcut at the corner. 

However, the agreement isn't perfect. The exact functional form of the almost-step-functions are different. And for $\tau > \tau_*$ the approach to saturation is also not exactly the same---though they both saturate exponentially, the analog model has a factor of two larger exponent and so saturates faster. However, as discussed in footnote~\ref{$k$-local}, the approach to saturation is not a robust feature of the epidemic model, and can be changed by changing $k$. \\

Let us now comment on the geometric interpretation of the size of the precursor. In \cite{Susskind:2014jwa} it was conjectured that it is related to the area $A(\tau)$ enclosed by the four geodesics in Fig.~\ref{oneprecursorsurface}. Here we make that statement more precise,
\begin{equation}
s[W(\tau)]=\frac{d\CC[W(\tau)]}{d\tau}=K\sin\frac{2A(\tau)}{K^2}.
\end{equation}
Though the length of geodesics keeps increasing, the area saturates at the scrambling time. This tells us that there exists a limiting curve that hugs the two Hamiltonian evolutions ($U(\tau)$ and $U^\dagger(\tau)$) when $\tau>\tau_*$. As time increases the blue geodesic in  Fig.~\ref{oneprecursorsurface} representing $W(\tau)$ will hug the limiting curve.

\subsection{Precursors: Multiple Perturbations}

In Sec.~\ref{subsec:multipleprecursors} we looked at the products of multiple precursors, and showed that so long as the individual precursors were separated by more than a scrambling time, the total switchback subtraction is  the  sum of the individual  subtractions at each switchback. In this subsection will see that the same is true for precursors in our analog model. 
\begin{figure}[H]
\begin{center}
\includegraphics[scale=.5]{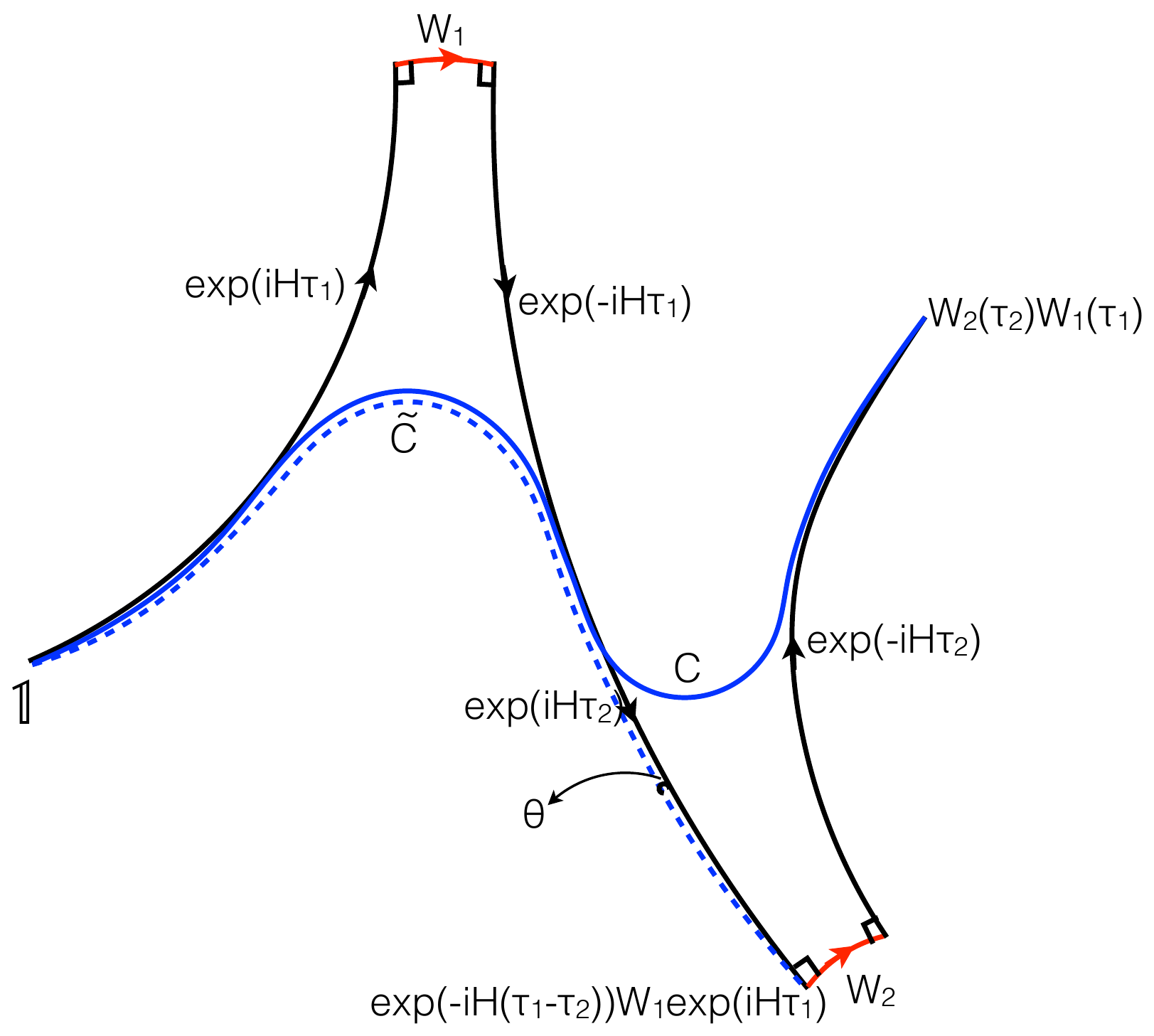}
\caption{Schematic representation of the product of two precursors on the negatively curved geometry. For large $|\tau_1|$, $|\tau_2|$, and $|\tau_1 -\tau_2|$, the length of the shortest geodesic connecting the start to the end is shorter than the sum of the five defining geodesic segments by an amount equal to two individual switchback subtractions.}
\label{twoprecursorsurface}
\end{center}
\end{figure}

The product of two precursors in the analog model is shown in Fig.~\ref{twoprecursorsurface}. The product is composed of $5$ geodesic segments: $e^{iH\tau_1}$, $W_1$, $e^{-iH(\tau_1-\tau_2)}$, $W_2$, and $e^{-iH\tau_2}$. By presumption the $W_i$ operators are orthogonal to the Hamiltonian, which in our analog model means that we choose the turns induced by $W$ to be through right-angles. The complexity of the precursor $\mathcal{C}[W_2(\tau_2)W_1(\tau_1)]$  corresponds to  the length of the shortest geodesic connecting the start to the end-point (shown in solid blue in Fig.~\ref{twoprecursorsurface}). 

The length of this shortest geodesic is calculated in Eq.~\ref{quadrupleturnanswer}. For $\CC[W_i] = 1 \ll K$, 
\begin{equation}
\cosh\frac{2\CC[W_2(\tau_2)W_1(\tau_1)]}{K} = 1 + \frac{2}{K^2} (\cosh [\tau_1] + \cosh [\tau_2])^2 + \frac{4}{K^4} \cosh [\tau_1] \cosh [\tau_2] \cosh \left[ \tau_1 + \tau_2 \right].
\end{equation}
This exhibits all the same phenomenology as the complexity of the multiple precursor operator in Sec.~\ref{subsec:multipleprecursors}. In particular, for $|\tau_1 - \tau_2 | \gg 2\tau_*$  the total shortcut from cutting the two corners is just the sum of the individual shortcuts at each corner. For example, for large $\tau_2$ but small $\tau_1$, the rate of change of complexity with $\tau_1$ grows like $e^{\tau_1}$. By contrast, when both $\tau_1$ and $\tau_2$ are large, 
\begin{align}
\mathcal{C}[W_2(\tau_2)W_1(\tau_1)]=\frac{K}{2}\Bigl(|\tau_1|+|\tau_1-\tau_2|+|\tau_2|-4\tau_*\Bigl) + \ldots, \label{eq:multipleswitchbackslength}
\end{align}
so that the geodesic shaves a distance $K \tau_*$ at each switchback. 

There is a geometric intuition for why distant switchback shortcuts simply add. Due to the divergence of geodesics in negatively curved spaces, the minimal geodesic must hew exponentially close to the $e^{i H \tau_1}$ geodesic until it gets within about a scrambling distance from the first corner. Having turned the first corner, within a scrambling distance it will once again hew to $e^{iH(\tau_2 - \tau_1)}$ until it gets within about a scrambling length of the second corner. The shortcuts at the two corners can only affect each other insofar as the middle segment of the geodesic is different from the $e^{iH(\tau_2 - \tau_1)}$ geodesic, but for $|\tau_1  - \tau_2| > 2 \tau_*$ that difference is exponentially small. 

As is discussed in Appendix~\ref{subsecfourrightangles}, this argument would apply equally had we made not two rights turns then two left turns, as in Fig.~\ref{twoprecursorsurface}, but rather four right turns. The argument would also apply with any number of switchbacks, and Eq.~\ref{multiswitchback} holds for a product of any number of precursors.

\subsection{Length and Action} \label{sec:lengthaction}
In our correspondence, the distance travelled by a particle moving along a geodesic corresponds to the complexity of the corresponding unitary operator. In this subsection, we examine the relation between the length of the particle's path and its on-shell action. We will show that they are linearly related. This suggests that we had a choice when interpreting  our analog model---to correspond to complexity, we could have chosen not geodesics length but on-shell action.

Let the line-element of the Riemannian space be 
\be 
dl^2 =g_{ij}dx^idx^j .
\label{R-metric}
\ee
The metric $g_{ij}$ has Euclidean signature and the coordinates do not include time, which as earlier we will denote by $\tau$. The action for the particle is
\be 
A = \int  \frac{m}{2} g_{ij}\dot{x}^i \dot{x}^j  \ d\tau.
\label{Action}
\ee
The action is different from the path length, which would be given by
\be 
l = \int \sqrt{ g_{ij}\dot{x}^i \dot{x}^j } d\tau.
\label{path-length}
\ee
The equation of motion following from the action Eq.~\ref{Action} is
\be 
\ddot{x}^i = g^{ik}\left[\frac{1}{2} \frac{\partial g_{mj}}{\partial x^k} -\frac{\partial g_{kj}}{\partial x^m}\right] \dot{x}^j\dot{x}^m;
\label{particle-equation}
\ee
particles that satisfy the equation of motion move along geodesics with constant velocity.
Along a geodesic, both the energy and the velocity of the particle  are conserved
\be 
E= \frac{m}{2} g_{ij}\dot{x}^i \dot{x}^j \ \ \ \  \& \  \  \ \ v =\sqrt{g_{ij}\dot{x}^i \dot{x}^j}  = \sqrt{\frac{2E}{m}}.
\ee
%
The path length Eq.~\ref{path-length} and the action Eq.~\ref{Action} are linearly related
\be  
A = \sqrt{\frac{mE}{2}}\ l.
\ee
The constant of proportionality can be set equal to one by choosing
\begin{equation} 
E = \frac{K}{2}  \ \ \ \ \& \ \ \ \ m = \frac{4}{K},
\end{equation}
so that the velocity of the particle is set to $K/2,$  which we have previously identified with  the rate of change of complexity. 

One point to note is that the expression for length in Eq.~\ref{path-length} is  invariant under reparametrization of the time variable $\tau,$ while the expression for action in Eq.~\ref{Action} is not. Equation~\ref{Action} has chosen a particular parametrization of the path, which is the conventions for time in Appendix A.  

In this paper we identified complexity with the minimum length of a trajectory; in this subsection we have shown that we could equally well have identified complexity with the minimum action\footnote{There is an argument that action is a better candidate to be the analog of complexity than length. This comes from considerations of the additivity properties of complexity under combining subsystems. Riemannian length adds in quadrature, whereas action, like complexity, is strictly additive. We thank Brian Swingle for explaining this point. These issues will be addressed in a forthcoming publication.} of any trajectory with velocity $K/2$ connecting the origin to the location of the particle. This is not the first time that complexity and action have been identified but the relation\footnote{Note that the action we are discussing is the action of the particle in the analog model, and not the action, as in \cite{Brown:2015bva,Brown:2015lvg}, of a  holographically dual gravitational system, nor the action of the strongly-coupled system itself---being strongly coupled, the system doesn't have a well-defined on-shell action.} to the results of \cite{Brown:2015bva,Brown:2015lvg} is not clear.


\sc
\section{Discussion} \label{sec:discussion}

Previous papers have exhibited a close match between quantum circuit  complexity and black hole geometry \cite{Susskind:2014rva,Stanford:2014jda,Roberts:2014isa,Brown:2015bva,Brown:2015lvg}. This paper exhibits a new correspondence between these two systems and a third---a non-relativistic particle moving on a negatively curved two-dimensional surface of large genus.

This  simple analog model  recovers many  properties of chaotic $k$-local Hamiltonian systems, as well as random circuits and black holes.  The negative curvature of the analog model  plays a key role, controlling  the exponential geodesic deviation, and therefore the quantum Lyapunov exponent  discussed in \cite{Maldacena:2015waa}. 

The quantum Lyapunov exponent can be shown to measure how rapidly the same initial quantum state deviates when subjected to two slightly different Hamiltonians. The classical Lyapunov exponent measures how rapidly two slightly different classical states deviate when subjected to the same Hamiltonian. Our correspondence is able to relate these two exponents by relating different fast-scrambling quantum Hamiltonians to different geodesics of the hyperbolic plane. \\

Moving a curvature distance on the hyperbolic plane corresponds to adding $K/2$ gates;  a unit distance corresponds to a single gate. On the hyperbolic plane we are able to resolve arbitrarily small distances. By contrast, and though it is a subject of active investigation, we do not yet have a fully satisfactory continuum definition of complexity for Hamiltonians such as those in Eq.~\ref{eq:$k$-local}. We speculate that perhaps the analog model could be used as a guide for constructing a continuum definition of complexity that is able to resolve complexities of a fraction of a gate.\\

A simple discrete version of our correspondence is visible already just by graphing the number of possible circuits as a function of depth. Consider the $2$-local random circuit model described in Fig.~1, and suppose there is exactly one kind of gate, which is a (non-symmetric) two-qubit gate. The  circuit with depth zero is the identity. After a single time-step, the number of different possible unitary transformations generated is generically
\begin{equation}
 \log \Bigl[ \# \textrm{ unitaries} \Bigl]  = \log \left[ \frac{K!}{(K/2)!} \right]  \approx  \frac{1}{2} K \log \Bigl[ \frac{2K }{e} \Bigl]  ;\label{eq:branches}
\end{equation} 
there is generically a different unitary for every distinct way of arranging the gates in the depth-one circuit. After each successive time-step the number of possible unitaries that may have been implemented by the circuit multiplies by almost as much again. It only fails to multiply insofar as there are collisions in which two different circuit designs give the same unitary, but as we argued in Sec.~\ref{sec2} these collisions are rare until the circuit is exponentially deep. Ignoring collisions, the graph forms a tree, and a tree is a discrete version of hyperbolic space.  

\begin{figure}[htbp] 
   \centering
   \includegraphics[height=1.4in]{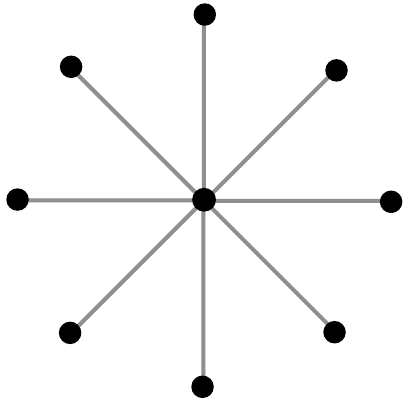} \hspace{3cm}
      \includegraphics[height=1.4in]{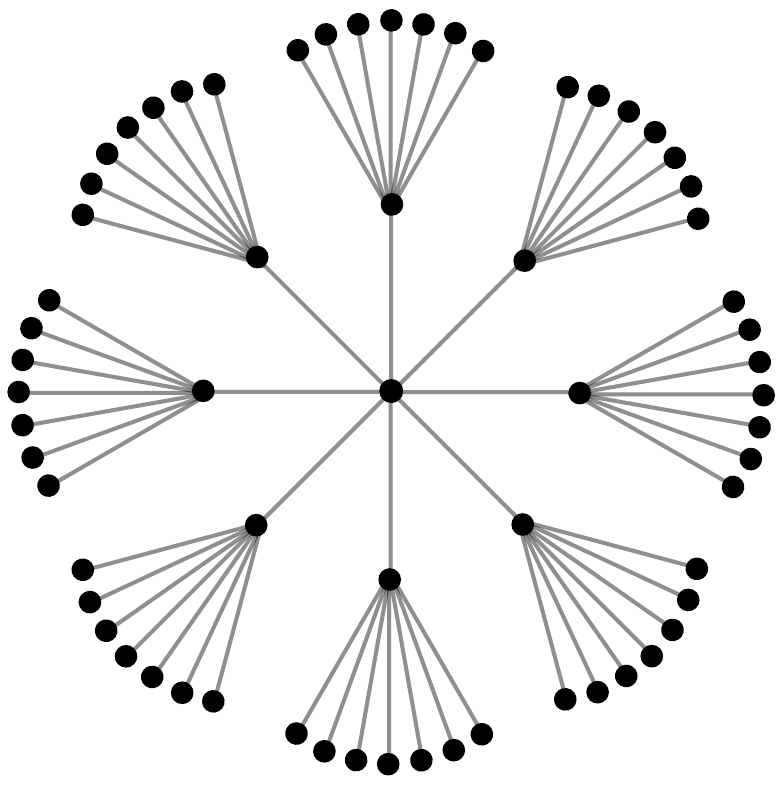} 
   \label{fig-treetwostep}
   \caption{The number of possible circuits grows exponentially with circuit depth. At depth zero, the circuit the identity. After a single time-step (on left), there are O($K^{K/2}$) possible unitaries that could have been created. After two time-steps (on right) there are O($K^K$). Almost all the possible unitaries reached after two steps are distinct, though there are rare coincidences. This forms a tree.  A Cayley tree is known to be a discrete model of a hyperbolic space.}
   \label{ffig-treeonestep}
\end{figure}

The correspondence that this toy discrete  version is capturing is that, on the one hand, the number of possible circuits grows exponentially with the depth, and, on the other hand, that the volume of a hyperbolic ball grows exponentially with radius. By examining the exponent in this hyperbolic growth, we can say something about the dimensionality of the hyperbolic space.

Throughout this paper, we considered geodesics on the two-dimensional hyperbolic plane $\mathbb{H}^2$. However, as shown in Appendix~\ref{sec:H2toHn}, essentially all our results about the match between complexity evolution and hyperbolic geodesics would continue to hold in $\mathbb{H}^D$. As discussed in Appendix~\ref{subsec:Pythagoras}, because of geodesic deviation on super-curvature scales, the behavior of geodesics in hyperbolic spaces are insensitive to the dimensionality. However,  the exponent with which volume increases with distance \emph{is} sensitive to the dimensionality. The volume enclosed within a ball of radius $r$ on a $\mathbb{H}^D$ of curvature length $K/2$ is 
\begin{equation}
\textrm{volume} \sim \exp \left[ \frac{2r (D -1)}{K} \right] .
\end{equation}
This grows more slowly than the rate at which the number of quantum circuits increases with depth in Eq.~\ref{eq:branches} unless the dimensionality of the hyperbolic space is  $D \sim \frac{1}{2} K \log K$.  

To capture the growth of complexity and the behavior of precursors it suffices to consider a hyperbolic space to any dimensionality---we have considered the simplest case of $D=2$. If we wish to also capture the rate at which the number of possible unitaries grows as a function of complexity, we must move to a higher-dimensional hyperbolic space.\footnote{Using the two-dimensional hyperbolic plane means that the typical point is at a radius $\Delta C \sim K$ less than the maximum radius, as in Eq.~\ref{eq:submaximalcomplexity}, corresponding to a complexity about $K$ below the maximum complexity. By instead adopting a higher-dimensional hyperbolic space, the typical complexity becomes much closer to the maximal complexity---for $D = \frac{1}{2} K \log K$ it is closer than a single gate.}    \\

We don't entirely understand why the parallel between our analog model and the evolution of quantum complexity works quite so well, but we suspect that the reason is an underlying relation with Nielsen's complexity geometry \cite{2007quant.ph..1004D}. Indeed, it was already pointed out in \cite{2007quant.ph..1004D} that the average sectional curvature of the complexity geometry is negative. The connection of our work with Nielsen's is explored in Sec.~\ref{sec:relation-nielsen-geometry}.

\subsection{A Triangle of Systems}

The three systems---black hole interior, quantum circuit, and analog model---form a triangle of ideas. The sides of the triangle are the relationships between pairs of these ideas. Here is how we understand those sides:

\bi 
\item Black hole---quantum circuit. This connection was proposed in \cite{Susskind:2014rva}. It seems reasonable to expect that the dynamics of a black hole can be represented as a Hamiltonian system of $K$ qubits with $K$ being approximately the entropy. A black hole is therefore a quantum circuit. However, the interesting connection is that the growth of complexity of the circuit is holographically dual to the growth of geometry behind the horizon. This strongly suggests that the emergence of space inside black holes is encoded in the growth of complexity. This side of the triangle is not new, but certainly not fully understood.

\item Quantum circuit---analog model. This is the connection that has been proposed in this paper. It seems likely  that this side of the triangle involves Nielsen's complexity geometry. Complexity geometry and the analog model share similar growth patterns: volume grows exponentially with distance. This, and the negative sectional curvatures of complexity geometry, are the key features that makes them similar.

\item  Analog model---black hole. For sub-exponential times 
both the analog model and the Einstein-Rosen bridge are described by classical geometry, but of quite different types. One is a fixed hyperbolic Euclidean signature space; the other is a time-dependent Lorentzian wormhole geometry.  It should be possible to bridge the gap between these two geometric systems in purely geometric terms without any quantum or information theoretic considerations, although at the moment we don't know how.

\ei

\section{Relationship to Nielsen Complexity Geometry}
\label{sec:relation-nielsen-geometry}

In this paper we have presented an analog model for the evolution of quantum complexity. This model was originally motivated by Nielsen's complexity geometry model \cite{2007quant.ph..1004D}. However the relationship between the models is not entirely clear---it's certainly not the case that we mathematically derived our model from Nielsen's.  In this section we present some thoughts about the connection between the two. 

In Nielsen's setup, we may regard the time evolution operator $U(\tau)$ as the motion of a classical particle on the space of unitary operators $SU(2^K)$ equipped with an unusual metric. The complexity of $U$ is related to the minimal geodesic distance between $U$ and the identity operator.  This is a relatively simple idea but  Nielsen's complexity geometry is very complicated and difficult to analyze. 

We can define a ``grand" analog model by considering a non-relativistic particle moving on the full complexity geometry, i.e., on $\suk$ equipped with the complexity metric \cite{2007quant.ph..1004D}
\be 
dl^2 = G_{MN}dX^M dX^N,
\ee
where the coordinates $X^M$ label points of $\suk.$
The usual metric on $\suk$ is invariant under both left and right multiplication by unitary operators; it has $\suk_L \times \suk_R$ invariance. The complexity metric, $G_{MN}$ has only right multiplication  invariance---this means that it is homogeneous, but unlike the standard metric not isotropic.  

The configuration space  (in the sense of classical mechanics) of the grand analog model is $\suk,$ but the phase space contains momenta as well as coordinates. The momenta are given in terms of the complexity metric by
\be 
P_M =  G_{MN}\dot{X}^N.
\ee

The equations of motion analogous to Eq.~\ref{particle-equation} are second order, and a trajectory is determined not only by an initial point $X$ but also the initial momenta. The momenta are in one-to-one correspondence with the generators of $\suk,$ and the allowable initial conditions require the momenta to correspond to  $k$-local generators, i.e.,  generators constructed from no more than $k$ Pauli operators.

On the other hand the motion of $U$ defined by the quantum Hamiltonian is determined  by the first order Schr\"odinger equation
\be 
\dot{U}=-i H U,
\ee
which means that for a given Hamiltonian only a single trajectory passes through  any $U.$

The resolution of this mismatch is that  the grand analog model does not describe a single quantum Hamiltonian; it describes all $k$-local Hamiltonians. In other words it describes all systems of the type in Eq.~\ref{eq:$k$-local}. In the quantum theory the future evolution of $U$ is determined not only by the current value of $U$ but also the set of coefficients $J_{i_1, i_2,...,i_k.}$ One can show that these coefficients correspond to the components of the analog momenta in the $k$-local directions.\\

An interesting consequence of this observation is that averaging over the phase space of the grand analog model involves a quenched average over an ensemble of $J$-coefficients, just as in the SYK model \cite{KitaevModel,Sachdev:1992fk,Polchinski:2016xgd,Maldacena:2016hyu}.
The natural framework for  such averaging   is classical statistical mechanics. For now we just remark that the ensemble-averaged  complexity translates to the classical entropy of the grand analog model. This helps explain why the evolution of complexity closely resembles the classical evolution of entropy. The much longer time-scale is due to the exponentially large number of degrees of freedom of the analog model. We will come back to this theme in a subsequent paper.\\

\begin{figure}[H]
\begin{center}
\includegraphics[scale=.35]{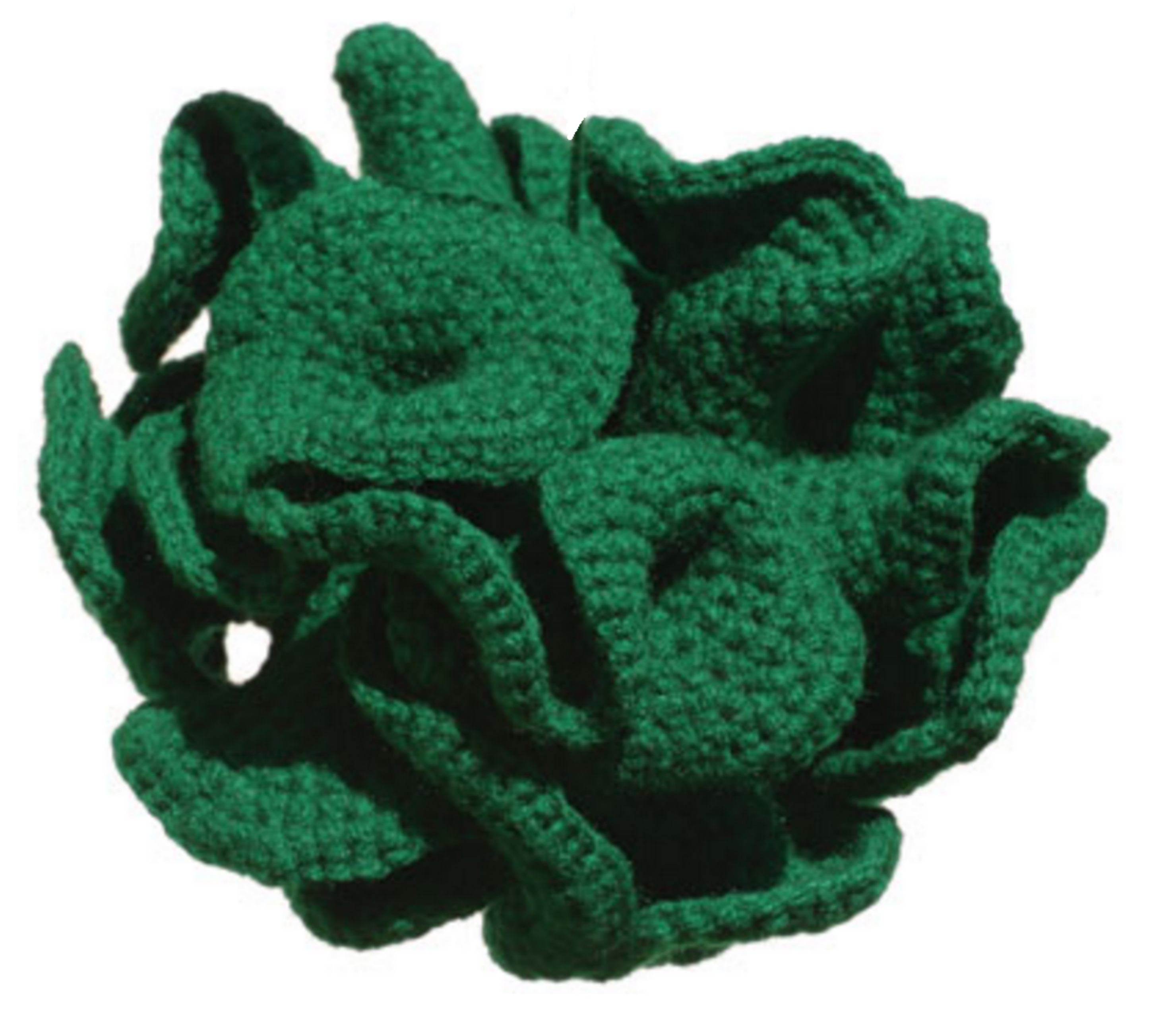}
\caption{We thank the mathematician and artist Daina Taimina for allowing us to use her crocheted model as an illustration of how space-filling the hyperbolic plane becomes when  embedded in flat space. The geometry of the surface is a slightly thickened hyperbolic plane. As it grows out from the center it will become densely packed at some radial distance from the starting point. The larger the dimension of the embedding space, the further the crocheting can proceed before becoming densely packed.  }
\label{crocheted}
\end{center}
\end{figure}

There are both similarities and differences between the geometry of the simple two-dimensional analog model and the complexity geometry of Nielsen. 
The most obvious difference is the dimensionality: complexity geometry has dimension $(4^K -1)$; the analog model has dimension $2.$ This difference seems very extreme, but there is a sense in which the hyperbolic plane mimics the properties of very high dimensional spaces. The dimension of a space reflects the growth of the volume as a function of linear dimension, i.e., the volume of a ball as a function of its radius. For sub-curvature distances the volume of the hyperbolic plane only grows quadratically with radius, but at super-curvature distances it grows faster than any power---it's like the dimensionality is infinite. Figure \ref{crocheted} is a photograph of an actual physical model of the hyperbolic plane embedded in 3-dimensional space. One sees how space-filling it becomes. 

\bn 
Let's consider the crocheted model embedded in $\mathbb{R}^n.$
Suppose the crocheting continues outward from the center for distance (in the surface) $L.$ The linear size in $\mathbb{R}^n$ will be less then $L$ but the volume of wool will be $ e^L \Delta$, where $\Delta$ is the thickness of the crocheted surface. The crocheting obviously cannot continue past the point where $L^n  =  e^L \Delta^{n-2}$ without becoming densely packed. The larger $n$, the further the crocheting can proceed; roughly, the maximum $L$ scales like $n.$  Now suppose $n$ is the dimension of $\suk,$ namely $n\sim 4^K.$ In that case the maximum linear dimension in the surface will be $\sim 4^K$ and the number of stitches will be of order $e^{4^K},$ in agreement with our cut off procedure.

A second big difference  is that complexity geometry is homogeneous whereas $\CH_g$ is not. One can pick any point in complexity geometry and transform it to the identity operator by a symmetry of the space; the minimal geodesic distance between points $U$ and $V$ is the \it relative \rm complexity of the corresponding operators. 
In the analog model, while $\mathbb{H}^2$ is homogeneous, the compactification renders it inhomogeneous---the topological identifications break the symmetry. This shows that we cannot   identify in a continuous manner the hyperbolic plane with the full Nielsen complexity geometry---it's not that points of $\CH_g$ represent individual operators in $\suk$ in a smooth way.  Instead, it seems that the region of the hyperbolic plane within $r\leq r_0$ represents the entire collection of operators with complexity $\CC \leq K r_0/2. $

The relationship between complexity geometry and the analog  model seems to be something along the following lines: begin by breaking the symmetry of complexity geometry by picking an arbitrary reference point (the crocheter's  starting point) and a 2-dimensional section, and then identify the reference point with the origin of $\CH_g.$  The area of the portion of $\CH_g$ less than a distance $L$ from the origin represents the collection of operators with relative complexity less than $L.$ For each reference point and each section there is a new set of $\CH_g$-surfaces; since each point of complexity geometry has its own identical copy of $\CH_g,$ complexity geometry is homogeneous but $\CH_g$ is not.

We hope to return to the relationship between Nielsen's complexity geometry and the analog model in future work.

\section*{Acknowledgements}

We thank Hrant Gharibyan, Patrick Hayden,  Sepehr Nezami,  Dan Roberts,  Steve Shenker,  and Douglas Stanford for discussions concerning some of the materials in this paper. 
This work was supported in
part by National Science Foundation grant 0756174.

\appendix
\section {Convention about Units} \label{appendixaboutunits}

\begin{itemize}
\item Throughout this paper we have been using a time variable, $\tau$, that treats complexity itself as a clock: the units of time are defined by the requirement that early on the rate of change of complexity is  $K/2$. 

\item For a typical qubit state, the effective temperature is infinite and the entropy $S$ is proportional to the number of qubits $K$; thus we identify the rate of change of complexity with the course-grained entropy of a typical state.  
 On the other hand, there are reasons to believe \cite{lloyd2000ultimate,Brown:2015bva} that for generic systems the early rate of change of complexity  is proportional to the internal energy. This means that our units of time must be such that the energy is equal to the entropy.

\item We note that even though the temperature of a random qubit state is infinite, the scrambling time does not go to zero. This is because the spectrum is bounded from above, so infinite temperature does not mean infinite energy per qubit. 

\item 
In the context of black holes, the time $\tau$ in this paper is not conventional Schwarzschild time, but is instead dimensionless Rindler time (i.e. the boost angle) 
\begin{equation}
\tau=\frac{2\pi}{\beta}\times \text{Schwarzschild time}.
\end{equation}
In these units, the energy is the entropy (divided by $2 \pi$). 

\item For the type of random circuit  discussed in Sec.~\ref{fastscramblers}, as well as for the complexity geometry, time $\tau$ is the depth of the circuit, i.e., number of parallel computing time-steps. At each time-step, $\frac{K}{2}$ gates act and the complexity generically increases by $\frac{K}{2}$; this is why the velocity of the analog particle is taken to be $\frac{K}{2}$.

\item
With this normalization of time, the limiting quantum Lyapunov exponent of \cite{Maldacena:2015waa} is $1$, and the corresponding scrambling time is $\tau_* = \log{K}.$ We expect   generic local systems (of the type for example studied in the SYK model \cite{KitaevModel}) will be close to saturating this bound. Thus the Lyapunov exponent is O(1); for simplicity in the paper we have put it exactly equal to one.

\item
The evolution $U(\tau)$ on Nielsen's geometry is expected to be chaotic \cite{2007quant.ph..1004D}. The motion of particles on a compact negatively curved surface is also chaotic. The classical Lyapunov exponent is determined by the sectional curvature from the equation of geodesic deviation. It is to be identified with the quantum Lyapunov exponent of the quantum circuit.

\end{itemize}

\section{Lengths of Geodesics in Hyperbolic Space} \label{appendixhyperbolicspace}
In this appendix, we review the geometry of hyperbolic space. In Sec.~\ref{subsec:Pythagoras} we explore the sense in which hyperbolic space has the standard $L_2$ norm at short distances but an $L_1$ norm at long distances. In Sec.~\ref{subsec:embedding} we derive the geometric results in Sec.~\ref{subsec:Pythagoras} and in the rest of the paper. Throughout this appendix we will consider hyperbolic spaces with unit radius of curvature---straightforward dimensional analysis can be used to translate this result to any other radius of curvature. 

\subsection{Pythagoras in Hyperbolic Space} \label{subsec:Pythagoras}

\subsubsection{$L_2$-norm on Short Distances, $L_1$-norm at Long Distances}
Consider an equal-sided right-angled triangle on a unit-sized hyperbolic plane. As a function of the length of the other sides, $s_1$, the length of the hypotenuse is (see Sec.~\ref{subsec:embedding}) 
 \begin{equation}
 \cosh[s_\textrm{hyp}]  =  \cosh [s_1]^2 . \label{eq:pythagoras}
 \end{equation}
  \begin{figure}[htbp] 
    \centering
    \includegraphics[width=1.3in]{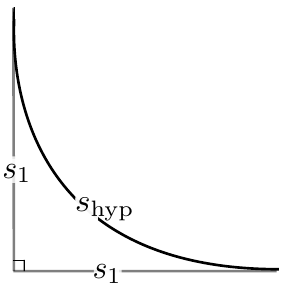} 
 \end{figure}
This demonstrates an essential property of hyperbolic space. On small scales, the hyperbolic plane has the usual L$_2$-norm, whereas on long scales it behaves as though it has an L$_1$-norm (the `taxicab geometry'):
\begin{equation}
\frac{ds_\textrm{hyp}}{ds_1} = \frac{2 \cosh[s_1]}{\sqrt{1 + \cosh[s_1]^2}}  = \Biggl\{
\begin{array}{cc}
\sqrt{2} &\textrm{ for } s_1 \ll 1 \, \,  \vspace{2mm} \\
2 & \textrm{ for } s_1 \gg 1 \, .
\end{array}
\end{equation}
Unlike in flat space, the hypotenuse closely tracks the right-angled sides, only substantively `cutting the corner' once it is within about a curvature length of the right angle. As a consequence, even for the largest triangles the total distance saved by the `shortcut' is only about a curvature length
 \begin{equation}
 s_\textrm{hyp} = 2s_1 - \log 2 + \textrm{O}(e^{-s_1}) . \label{thepythogorianshortcut}
 \end{equation}
 
\subsubsection{Shortcuts on the Hyperbolic Plane}
It is interesting to generalize this formula to the case where the angle is no longer a right angle and the two sides have different length.  
   \begin{figure}[htbp] 
   \centering
   \includegraphics[height=1.4in]{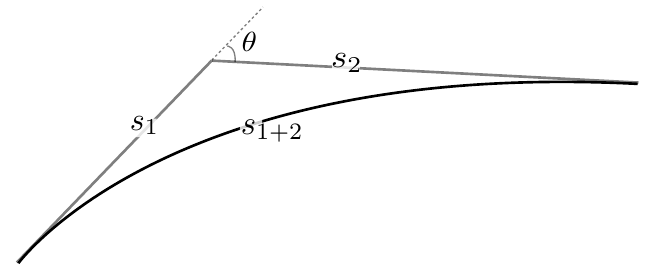} 
   \label{fig:cuttingcorner}
\end{figure}
In that case the answer is 
\begin{equation}
\cosh s_{1+2} = \cosh s_1 \cosh s_2 + \cos \theta \sinh s_1 \sinh s_2. \label{singleturnanswer}
\end{equation}
For $\theta =0$ (no turn) this gives $s_{1+2} = s_1 + s_2$. For $\theta = 90^\circ$ and $s_1 = s_2$ this gives Eq.~\ref{eq:pythagoras}. For $\theta = 180^\circ$ (about face) this gives $s_{1+2} = |s_1 - s_2|$. For $s_1, s_2 \gg 1$ this gives 
\begin{equation}
s_{1 + 2} = s_1 + s_2  + \log \cos^2 \frac{\theta}{2} + \dots. \label{latetimesoneturn}
\end{equation} This means that  smoothing the sharp corner shortens the path by about a curvature length, $\Delta s \sim -1$, unless $\theta$ is exponentially close to $180^\circ$, in which case $\Delta s \sim 2 \log |\frac{\theta - 180^{\circ}}{2}|$. 

Roughly speaking, the shortcutting geodesic hugs the original geodesics until the two original geodesics are a curvature length apart. For intermediate $\theta$ the two original geodesics are a curvature length apart when the distance to the corner is about a curvature length. For tiny $\pi - \theta$ the two geodesics are much closer together, and remain less than a curvature length apart until a scrambling distance from the corner. 

\subsubsection{Two Right-Angled Turns}
Now consider making two turns. Specifically, you walk along a geodesic for a distance $s_2$, turn right, walk another  $2s_1$, turn right again, and walk a further $s_2$.
   \begin{figure}[htbp] 
    \centering
    \includegraphics[width=3in]{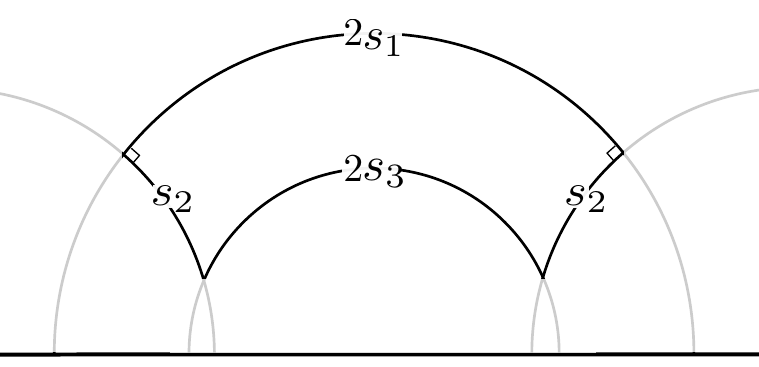} 
 \end{figure}
  The geodesic distance back to the starting point is 
 \begin{eqnarray}
 \cosh[2s_3] & = & \cosh [s_1]^2 + \sinh[s_1]^2 \cosh [2s_2]   \label{eq:theanswer} \Bo \\
 \rightarrow  \ \ \frac{ds_3}{ds_2} &=&  \frac{2 \sinh[s_1] \sinh [s_2]}{\cosh[s_3]}. \label{eq:rateofchange}
 \end{eqnarray}
For small $s_1$, the growth rate Eq.~\ref{eq:rateofchange} looks like 
\begin{figure}[htbp] 
   \centering
   \includegraphics[width=3in]{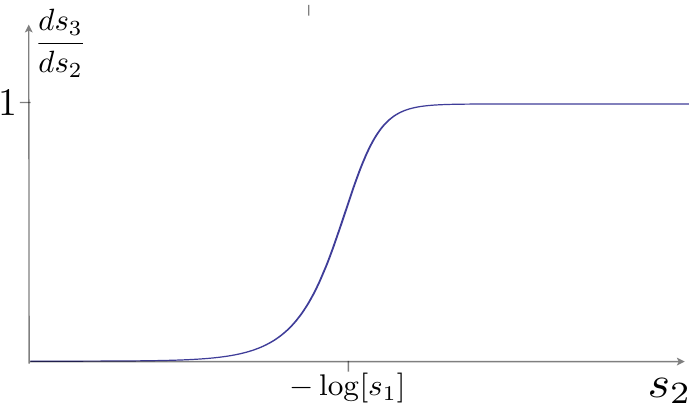} .
   \label{fig:growthatsmalls1}
\end{figure}

\noindent At early times ($s_2\,  \, \lsim - \log s_1 \, \,  \leftrightarrow \, \, s_3 \, \, \lsim  \, \, 1$), $s_3$ grows exponentially
\begin{equation}
\textrm{early times:} \ \ s_3 = s_1 \cosh s_2. 
\end{equation}
(This early-time formula follows directly from the equation of geodesic deviation.) \\

\noindent At late times ($s_2\,  \, \gsim - \log s_1 \, \, \leftrightarrow \, \, s_3 \, \, \gsim \, \, 1$), $s_3$ grows linearly
\begin{equation}
\textrm{late times:} \ \ s_3 = s_2 + \log [\sinh s_1] . \label{eq:latetimes}
\end{equation}
The switchback delay is $\log [\sinh s_1] $, which is about a scrambling distance. \\

\noindent Finally, notice that for large $s_1$ and $s_2$, Eq.~\ref{eq:theanswer} gives
\begin{equation}
2s_3 = 2 s_1 + 2s_2 - 2 \log 2 + \ldots.
\end{equation}
The distance shaved by the shortcut  is $2 \log 2$. Thus the distance shaved by shortcutting two well-separated turns is double the distance shaved by shortcutting a single turn in Eq.~\ref{thepythogorianshortcut}; one might be tempted to say that two well-separated turns do not `interact'. We'll see this again in the example of the next subsection. 

\subsubsection{Four Right-Angled Turns} \label{subsecfourrightangles}

Walk a distance $s_1$, turn right, walk  $s_2$, turn right, walk $s_3$, turn left, walk $s_4$, turn left, walk $s_5$. The geodesic distance back to the start is 
\begin{eqnarray}
 \cosh s_{1+2+3+ 4+5} & =&  \cosh s_1 \cosh s_2  \cosh s_3  \cosh s_4  \cosh s_5 \label{quadrupleturnanswer} \\
& & \ - \sinh s_1 \sinh s_3 \cosh s_4 \cosh s_5 - \cosh s_1 \cosh s_2 \sinh s_3 \sinh s_5  \nonumber \\
& & \ \ + \cosh s_1 \sinh s_2 \sinh s_4 \cosh s_5 + \sinh s_1 \cosh s_3 \sinh s_5 . \nonumber 
\end{eqnarray}
 \begin{figure}[htbp] 
   \centering
   \includegraphics[height=1.7in]{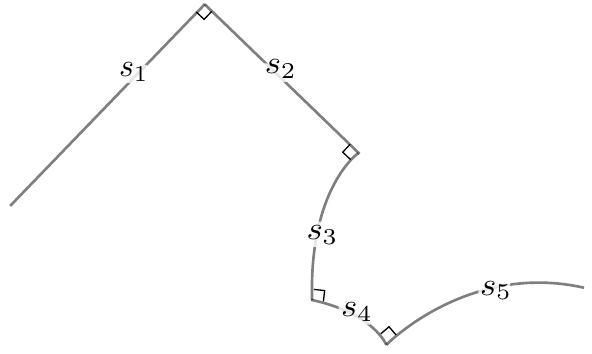} 
   \label{fig-fourturns}
\end{figure}
\noindent For $s_2, s_4 \ll 1 \ll s_1, s_3,s_5$ this gives
\begin{equation} 
s_{1+2+3+4 + 5} = s_1 + s_3 + s_5 + 2 \log \frac{s_2}{2} + 2 \log \frac{s_4}{2}.
\end{equation}
The shortcut thus shaves a scrambling distance at each of the two about-turns. \\

\noindent The four successive turns were first right, then right, then left, and then left. Had we instead made four successive right turns, this would correspond to using negative $s_4$ in Eq.~\ref{quadrupleturnanswer}. Since only the fourth term is sensitive to the sign of $s_4$, and since that term is exponentially smaller than the others when $s_3$ is large, four successive right turns would lead to a shorter $s_{1+2+3+4+5}$, but only by an exponentially small amount.

\subsubsection{From $\mathbb{H}^2$ to $\mathbb{H}^D$} \label{sec:H2toHn}

So far our results have been for the hyperbolic plane, $\mathbb{H}^2$. However, the crucial  feature that hyperbolic space looks like it has an $L_2$-norm on short distances and an $L_1$-norm on supercurvature scales persists in any dimension. For example, consider $\mathbb{H}^3$. You walk for $s_1$, turn right, walk another $s_2$, turn right, turn by an angle $\theta$ into the third dimension, and then walk a further $s_3$.  The geodesic distance back to where you started is 
 \begin{equation}
\cosh[ s_{1 + 2 + 3}] =  \cosh[s_1] \cosh[s_2] \cosh[s_3] - \sinh [s_1] \sinh[s_3] \cos \theta.
 \end{equation}
 For large $s_2$, the term that depends on $\theta$ is exponentially suppressed. 

Indeed, consider $\mathbb{H}^D$. You walk for $s_1$, turn 90$^\circ$, walk another $s_2$, turn 90$^\circ$ orthogonal to your first turn, walk another $s_3$, turn another 90$^\circ$ orthogonal to both your previous turns\ldots. Every time you turn, you turn into a fresh dimension of the $\mathbb{H}^D$. The geodesic distance back to where you started is given simply by
\begin{equation}
\cosh[ s_{1 + 2 + \ldots + D}] =  \cosh[s_1] \cosh[s_2] \ldots \cosh[s_D].
 \end{equation}
 This confirms that essentially all our results would carry over from $\mathbb{H}^2$ to $\mathbb{H}^D$.

\subsection{Derivation from Embedding Space} \label{subsec:embedding}
The easiest way to calculate distances is to embed hyperbolic space into a Minkowski space of one more dimension. Let's  start with $\mathbb{H}^2$. Three-dimensional Minkowski space has metric 
\begin{equation}
ds^2 = -dT^2 + dX^2 + dY^2 .
\end{equation}
The hyperbolic plane is given  by restricting to points whose displacement from the origin, $\vec{v}$, is a  forward-directed timelike vector of unit length:
\begin{equation}
\vec{v} \cdot \vec{v}  \equiv -T^2 + X^2 + Y^2 = -1.
\end{equation}
This gives the unit hyperbolic plane, which has curvature length $1$,  Gaussian curvature $-1$, and Ricci curvature $\mathcal{R} = -2$. The 1 rotation and 2 boost symmetries that leave the origin fixed in Minkowski$_{2+1}$  become the 3 symmetries of $\mathbb{H}^2$. The normalized normal to the hyperbolic place is $\vec{n} = \vec{v}$. We can recover standard radial coordinates $ds^2 = dr^2 + \sinh^2 \hspace{-1pt} r \hspace{1pt} d \theta^2$ by writing 
\begin{equation}
v^a = \left( \begin{array}{c}
T \\
X \\
Y 
\end{array} \right) 
= 
\left( \begin{array}{c}
\cosh r \\
\sinh r \cos \theta \\
\sinh r \sin \theta 
\end{array} \right)  ,
\end{equation}
which manifestly has $v^a v_a = -1$. The distance between  the points $\vec{u}$ and $\vec{v}$ is their relative rapidity 
\begin{eqnarray}
\cosh[ s_{{u} {v}}] &=& - \vec{u} \cdot \vec{v}  \label{cosh} \\
\sinh[ s_{{u} {v}}] &=& |\vec{u} \times \vec{v}| . \label{sinh}
\end{eqnarray}
Geodesics are given by intersecting  $\mathbb{H}^2$ with a plane through the origin. The geodesic through $\vec{u}$ and $\vec{v}$ has normalized normal
\begin{equation}
\hat{n}_{uv} = \frac{ \vec{u} \times \vec{v}}{| \vec{u} \times \vec{v}|}. 
\end{equation}
The locus of points on the geodesic is those $\vec{w}$ for which $\vec{w} \cdot \vec{w} = - 1$ and $\vec{w} \cdot \vec{n} = 0$. \\

\noindent The angle between the planes $\vec{n}_1$ and $\vec{n}_2$, as measured from the plane $\vec{n}_3$, is given by 
\begin{equation}
\sin \theta = \hat{n}_1 \cdot \hat{n}_2 \times \hat{n}_3.
\end{equation}
Therefore the angle between the geodesic from $\vec{u}$ to $\vec{v}$ and the geodesic from $\vec{v}$ to $\vec{w}$ is 
\begin{eqnarray}
\sin \theta  &= &\frac{ \vec{u} \times \vec{v}}{| \vec{u} \times \vec{v}|} \cdot \vec{v} \times \frac{ \vec{v} \times \vec{w}}{| \vec{v} \times \vec{w}|} = \frac{ \vec{u} \cdot \vec{v} \times \vec{w}}{| \vec{u} \times \vec{v}| | \vec{v} \times \vec{w}|} 
\end{eqnarray}\\
%
 Using embedding coordinates, let's give a general prescription for calculating the geodesic length between the start and endpoints of any paths of the form `proceed for a distance $s_1$, then turn through an angle $\theta_{12}$, then proceed for a distance $s_2$, then turn through $\theta_{23}$, then \ldots''. 

Without loss of generality, we will begin in coordinates such that the starting point is at the origin, ${u}^a = (1,0,0)$, and such that we set off in the $X$-direction. Then we'll walk along the path, changing coordinates as we go to keep ourselves at the origin, and keep ourselves pointing in the $X$-direction, so that by the time we arrive we'll in be coordinates such that the final point is at the origin ${v}^a = (1,0,0)$.  Let's begin. After we've walked a distance $s_1$, in the new coordinates  $\vec{u}$ is 
\begin{equation}
u^a = \left( \begin{array}{ccc}
\cosh s_1 & \sinh s_1 & 0 \\
\sinh s_1 & \cosh s_1 & 0 \\
0 & 0 & 1 
\end{array} \right) 
\left( \begin{array}{c}
1 \\
0 \\
0 
\end{array} \right)  .
\end{equation}
We then turn left by $\theta_{12}$, and then proceed by a further $s_2$  in the new $X$-direction. After doing this $n$ times,  the geodesic distance from $\vec{u}$ to $\vec{v}$ can be calculated using Eq.~\ref{cosh} as 
\begin{eqnarray}
\cosh s_{1 + 2 + \ldots + n} = - v_a u^a &=&   \begin{array}{ccc}
\Bigl( 1 & 0 & 0 \Bigl) \\ 
&& \\ 
&&  \end{array}
\left( \begin{array}{ccc}
\cosh s_n & \sinh s_n & 0 \\
\sinh s_n & \cosh s_n & 0 \\
0 & 0 & 1
\end{array} \right)  \ldots  \\
&& \ \ \  \ldots
\left( \begin{array}{ccc}
1 &0 & 0 \\
0 & \cos \theta_{12} & \sin \theta_{12} \\
0 & - \sin \theta_{12}  & \cos \theta_{12}  
\end{array} \right) \left( \begin{array}{ccc}
\cosh s_1 & \sinh s_1 & 0 \\
\sinh s_1 & \cosh s_1 & 0 \\
0 & 0 & 1 
\end{array} \right) \left( \begin{array}{c}
1 \\
0 \\
0 
\end{array} \right)  \nonumber .
\end{eqnarray}
As special cases of this general formula, we recover  Eqs.~\ref{eq:pythagoras}, \ref{singleturnanswer}, \ref{eq:theanswer} \& \ref{quadrupleturnanswer}, and thence Eqs.~\ref{oneprecursor} \& \ref{eq:multipleswitchbackslength}.

\bibliographystyle{ut}

\bibliography{reference}

\end{document}